\documentclass[journal=jacsat,manuscript=article]{achemso}

\usepackage{chemformula} 
\usepackage[T1]{fontenc} 
\usepackage{graphicx} 
\usepackage{amsmath}
\usepackage{color}
\usepackage{multirow}
\usepackage{amssymb}
\usepackage{dcolumn}%
\usepackage{bm}%
\usepackage{ctable}
\usepackage{tabularx}
\usepackage{graphicx}
\newcommand{\marktext}[1]{\textcolor{black}{#1}}

\title[\texttt{achemso}]{Spatial Dimensionality Dependence of  Heterogeneity, Breakdown of the Stokes-Einstein Relation and Fragility of a Model Glass-Forming Liquid}

\author{Monoj Adhikari}
\affiliation{Jawaharlal Nehru Centre for Advanced Scientific Research, Jakkuru Campus, 560064 Bengaluru, India}

\author{Smarajit Karmakar}
\affiliation{TIFR Center for Interdisciplinary Science, Tata Institute of Fundamental Research,36/P Gopanpally Village, Serilingampally Mandal, RR District, Hyderabad, 500075, Telangana, India,}
\author{Srikanth Sastry}
\email{sastry@jncasr.ac.in}
\affiliation{Jawaharlal Nehru Centre for Advanced Scientific Research, Jakkuru Campus, 560064, Bengaluru, India}

\begin{document}

\begin{abstract}
We investigate the heterogeneity of dynamics, the breakdown of the Stokes-Einstein relation and fragility in a model glass forming liquid, a binary mixture of soft spheres with a harmonic interaction potential, for spatial dimensions from $3$ to $8$. Dynamical heterogeneity is quantified through the dynamical susceptibility $\chi_4$, and the non-Gaussian parameter $\alpha_2$. We find that the fragility, the degree of breakdown of the Stokes-Einstein relation, as well as heterogeneity of dynamics, decrease with increasing spatial dimensionality. We briefly describe the dependence of fragility on density, and use it to resolve an apparent inconsistency with previous results. 
\end{abstract}

\maketitle

\section{Introduction}
When a glass-forming liquid is cooled to low temperatures, its dynamics becomes heterogeneous; spatially correlated clusters of particles move faster or slower than the average. The growth of dynamic heterogeneity (DH) and its associated dynamic length scales with the lowering of temperature have been considered a hallmark of relaxation dynamics in glass forming liquids \cite{sillescu1999heterogeneity,ediger2000spatially,berthier2011dynamical,kob1997dynamical,perera1998origin,yamamoto1998heterogeneous,donati2002theory,flenner2014universal}. The role of such heterogeneities in the complex relaxation dynamics in glass forming liquids, and growing length scales governing these relaxation processes have been widely investigated \cite{karmakar2014growing,karmakar2016length}. An extensively studied phenomenon, which has been analysed in the context of dynamical heterogeneity is the breakdown of the Stokes-Einstein relation (SER) ${D \eta \over T} = constant$ (or equivalently, ${D \tau_{\alpha}} = constant$) where $D$ is the diffusion coefficient, $\eta$ is the shear viscosity, $T$ is the temperature, and $\tau_{\alpha}$ is the structural relaxation time \cite{Rossler1990,fujara1992translational,thirumalai1993activated,stillinger1994translation,tarjus1995breakdown,cicerone1996enhanced,berthier2004length,berthier2004time,Jung2004,kim2005breakdown,kumar2006,becker2006fractional,chong2008connections,chong2009coupling,sengupta2013breakdown,sengupta2014distribution,charbonneau2013dimensional,charbonneau2014hopping,parmar2017length}. Although the origin of such a breakdown in hopping dynamics have also been investigated \cite{chong2008connections,charbonneau2014hopping,parmar2017length}, significant evidence links the breakdown of the Stokes-Einstein relation with length scales over which dynamics is heterogeneous\cite{chong2008connections,nandi2019continuous,parmar2017length}. It has also been suggested that fragility, which quantifies the degree of non-Arrhenius increase of relaxation times upon lowering temperature, is also related to heterogeneous dynamics. 
B\"ohmer \emph{et al.} \cite{bohmer1993nonexponential} investigated the correlation between the fragility and heterogeneity of dynamics, by compiling data for a large number of glass formers. Fragility was quantified by the fragility index $m$ which measures the steepness of rise of relaxation times at the glass transition, in an Angell plot\cite{angell1991relaxation}, wherein the logarithm of the relaxation time is plotted against inverse temperature scaled to the glass transition temperature ($T_g/T$). 
The KWW exponent, $\beta$, which characterises stretched exponential relaxation of density fluctuations, was considered  as a measure of the heterogeneity of dynamics.
Large values of fragility index were found to correspond to small values of $\beta$. Such a correlation between heterogeneity and fragility have been probed in several works (and also contested \cite{Dyre2009}), typically through consideration of the relationship between configurational entropy, fragility and cooperative length scales of dynamics \cite{douglas2006does,dudowicz2007,starr2013relationship,betancourt2013fragility}. Some of the issues involved have been addressed within the framework of the random first order transition (RFOT) theory \cite{kirkpatrick1987dynamics,kirkpatrick1988comparison,kirkpatrick1989scaling}, extensions of mode coupling theory \cite{biroli2006inhomogeneous},  recent exact results in the limit of infinite spatial dimensions\cite{charbonneau2017glass} and corresponding investigations of dynamics in variable dimensions \cite{Manacorda2020}. In particular, analyses of static and dynamic behaviour that may be expected in finite dimensions have led to identification of an upper critical dimension of $d_u = 8$ above which mean-field theories provide the correct description\cite{Biroli_2007,Franz18725}. Within the framework of the generalised entropy theory as well\cite{xu2016entropy}, $d = 8$ arises as a special dimension, above which an entropy vanishing transition does not exist at finite temperature.
In this context, it is of interest to understand how aspects of heterogeneous dynamics, the breakdown of the SER, and fragility depend on spatial dimensionality. Indeed, some studies have addressed such dependence \cite{eaves2009spatial,sengupta2013breakdown,charbonneau2013dimensional}.  
In particular, Charbonneau \emph{et al.} \cite{charbonneau2013dimensional} considered hard sphere fluids up to $10$ dimensions and showed evidence that the exponent $\omega$ in the relation $D \sim \tau_{\alpha}^{-1 + \omega}$ that quantifies the break down of the SER vanishes within numerical uncertainty above spatial dimension $d = 8$. 
However, this remains the only study that has explored the dimension dependence above $d = 4$. A similar study, for a model system with an interaction potential other than hard core interaction, which permits the study of both temperature and density dependent behaviour, is therefore desirable. We undertake such a study in the present work. 
\\
We investigate a model glass forming liquid consisting of a binary mixture of spheres interacting with a harmonic potential, in $3-8$ spatial dimensions. In the zero-temperature limit, this model has the limiting behaviour of the hard sphere model whose behaviour is controlled by density alone, while it exhibits behaviour of dense glass formers at high densities, at finite temperature. We perform computer simulations and investigate various measures of dynamical heterogeneity (DH) such as the non-Gaussian parameter, $\alpha_2(t)$, the dynamical susceptibility, $\chi_4$ as a function of time for a wide range of temperatures.  We compute the fragilities from the temperature dependence of the relaxation times, and further investigate the breakdown of the SER from a comparison of diffusion coefficients and relaxation times. We find a consistent variation of behaviour as the spatial dimension increases, wherein the fragility, extent of heterogeneity, and the degree of breakdown of the SER decrease with increasing spatial dimensionality, consistent with the approach to mean-field behaviour at $d = 8$. We briefly discuss the dependence on density of fragility and resolve an apparent inconsistency with previously published results which suggested an increase of fragility with increasing dimension while the degree of heterogeneity decreased. 

The rest of the paper is organized as follows: In section II, we describe the model and methods related to this study. In section III, we present our main results, and in section IV, we briefly discuss the density dependence of fragility and make a comparison with earlier work. Finally, we present a summary of results and conclusions in section V.\\

\section{Simulations details}

We investigate a $50:50$ binary mixture of particles that interact with a harmonic potential given by \cite{durian1995foam,berthier2009compressing}: 
\begin{eqnarray}
\nonumber
V_{\alpha \beta}(r) &=& \frac{\epsilon_{\alpha \beta}}{2}\left(1-\frac{r}{\sigma_{\alpha \beta}}\right)^{2}, \hspace{1.58cm} r_{\alpha \beta} \leq \sigma_{\alpha \beta}\\ 
&=&0 ,\hspace{3.94cm} r_{\alpha \beta } > \sigma_{\alpha \beta}
\end{eqnarray}
where $\alpha, \;  \beta$ $\in$ (A,B), indicates the type of particle. The two types of particle differ in their size, with $\sigma_{BB} =1.4 \sigma_{AA}$ (and the diameters are additive), with the interaction strengths being the same for all pairs. 
We present results for $3d$ - $8d$ fixing the density at $1.3\phi_J$, where $\phi_J$ is the jamming density. We have used $\phi_J$=$0.645 (3$d), $0.467 (4$d), $0.319(5$d), $0.209(6$d), $0.133 (7$d), $0.0821(8$d), using estimates by Charbonneau \emph{et al.} \cite{charbonneau2011glass}. 
In an accompanying study, we report our estimates of $\phi_J$, as well as a dynamical cross over density $\phi_0$. Our estimates of $\phi_J$ are close to the values used here to a high degree of accuracy. Thus, the volume fraction we use in our simulations are as follows: $0.8384$ ($3$d), $0.6071$ ($4$d), $0.4147$ ($5$d), $0.2717$ ($6$d),$0.1729$ ($7$d), $0.1067$ $(8$d).
The number density, $\rho$ is related to the volume fraction $\phi$ for the binary mixture in  the following way 
\begin{equation}
\phi =  \rho 2^{-d}\frac{\pi^{d/2}}{\Gamma(1+\frac{d}{2})}((c_A\sigma^d_{AA} + c_B\sigma^d_{BB}) 
\end{equation}   
where $\rho = N/V$, with $N$ being the number of particles, and $V$ the volume, and the fractions $c_A = c_B = 1/2$. The corresponding number densities are following: $0.8556$ ($3$d),$0.8132$ ($4$d), $0.7904$ ($5$d), $0.7891$ ($6$d), $0.8114$ ($7$d), $0.8554$ ($8$d).

The system size is fixed at $5000$ particles, which is large enough that the linear dimension $L$ is $> 2 \sigma_{BB}$ in all dimensions. Molecular dynamics (MD) simulations are performed in a cubic box with periodic boundary conditions in the constant number, volume, and temperature (NVT) ensemble. The integration time step was fixed at $dt=0.01$. Temperatures are kept constant using the  Brown and Clarke \cite{brown1984comparison} algorithm. The data, presented here, have run lengths of around $100\tau$ (where $\tau$ is the relaxation time, defined below). We present results that  are averaged over five independent samples. 
For results over a range of densities which we discuss in section IV, results are from 1-2 independent samples at densities other than those mentioned above.
We use reduced units with the small particle diameter, $\sigma_{AA}$, as the unit of length, $\epsilon_{AA}$ as the energy unit, and $\sqrt \frac{\sigma_{AA}^2 m_{AA}}{\epsilon_{AA}}$ as time unit, where $m_{AA}$ is the mass which is set to unity. 

\section{Results}
\subsection{Fragility in different dimensions} 

We quantify the microscopic dynamics by computing the overlap function, which is defined (for the $B$ particles) by: 
\begin{equation}
q(t) = \frac{1}{N_B}\sum_{i}^{N_B} w(|\textbf{r}_i(t_0)- \textbf{r}_i(t+t_0)|)  
\quad \mbox{where} \quad
w(x)=
\begin{cases}
1.0 & \text{if $x \le a$}\\
0 & \text{otherwise.}
\end{cases}\\ \nonumber
\end{equation}
 
\begin{figure}[htp]
\centering
\includegraphics[width=0.45\textwidth]{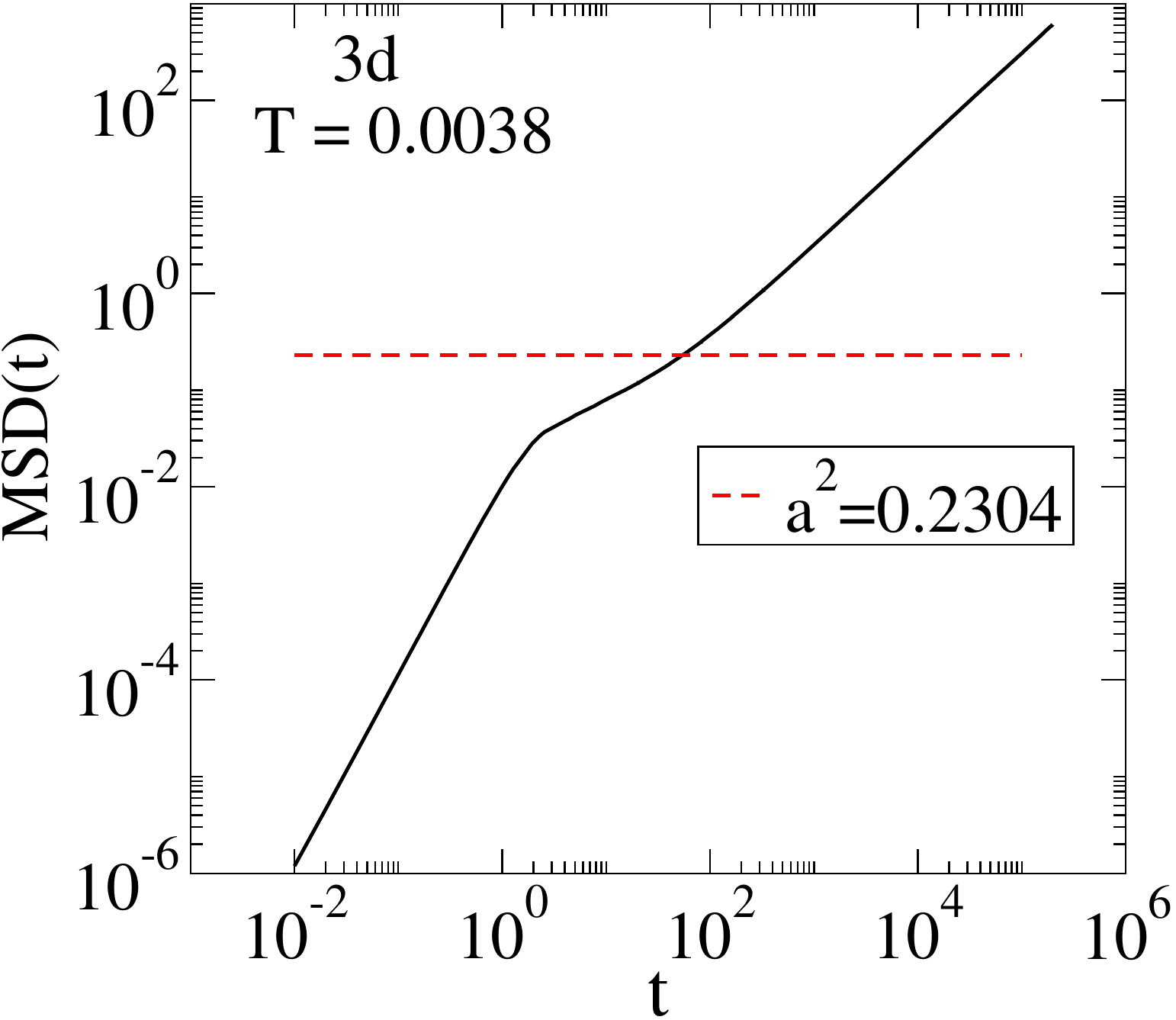}
\includegraphics[width=0.45\textwidth]{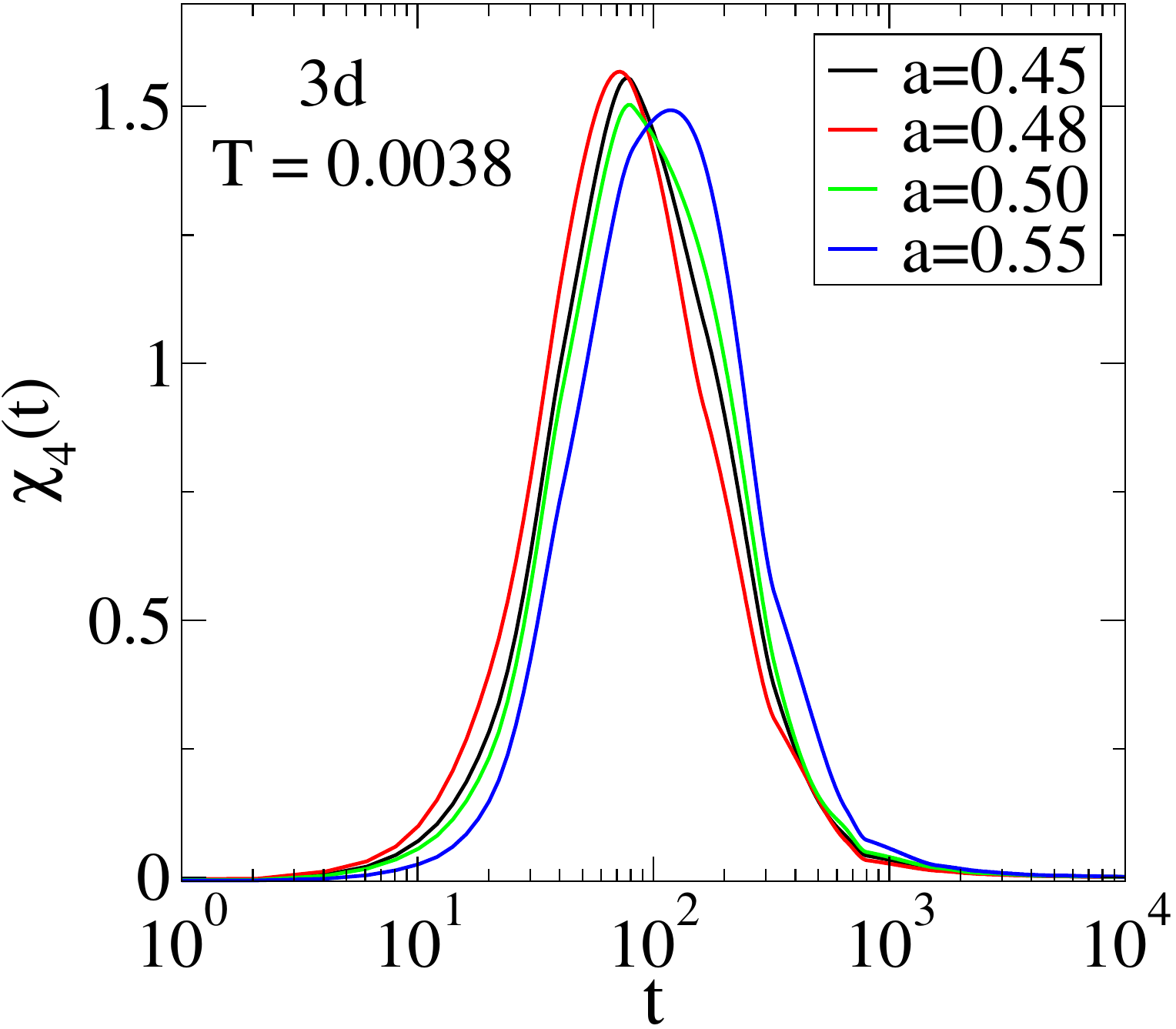}
\caption{Choice of the cut-off parameter {\it a} for calculating the overlap function, illustrated for $3$d, with $T=0.0038$ as the representative temperature. Left: We show the mean squared displacement (MSD) as a function of time. The horizontal red line corresponds to $a^2=0.2304$. Right: $\chi_4(t)$ is shown as a function of time for different {\it a} values. The choice $a = 0.48$ leads to the maximum peak value of $\chi_4$.}
\label{cutoff-choose}
\end{figure}

Here {\it a} is the cut-off within which particle positions are treated as indistinguishable\cite{lavcevic2003spatially}. We choose the parameter {\it a} in such a way that $a^2$ is close to the plateau value of the mean squared displacement (MSD). The choice of {\it a} is further refined by considering the behaviour of $\chi_4$ (defined below), to identify the value of {\it a} for which the peak value of $\chi_4$ is maximum with the choice of {\it a}. Fig.~\ref{cutoff-choose} illustrates the choice for three dimensions (3d), and the {\it a} value for other dimensions is chosen by a similar procedure. We choose the parameter values $a =0.48, 0.50, 0.55, 0.60, 0.75$ ,and $0.80$ for $3$d, $4$d, $5$d, $6$d, $7$d, and $8$d, respectively.

\begin{figure}[htp]
\centering
\includegraphics[width=0.50\textwidth]{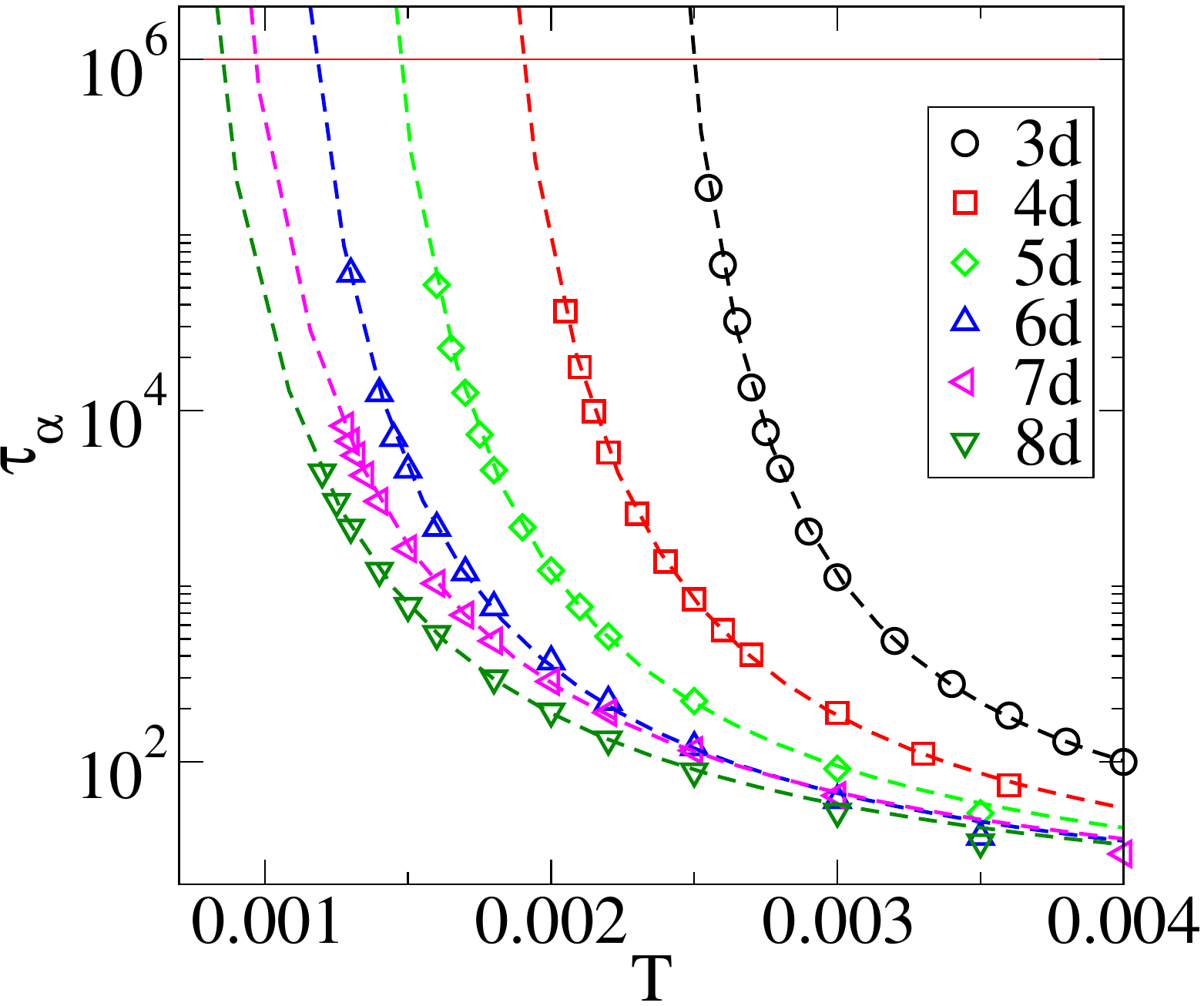}
\includegraphics[width=0.46\textwidth]{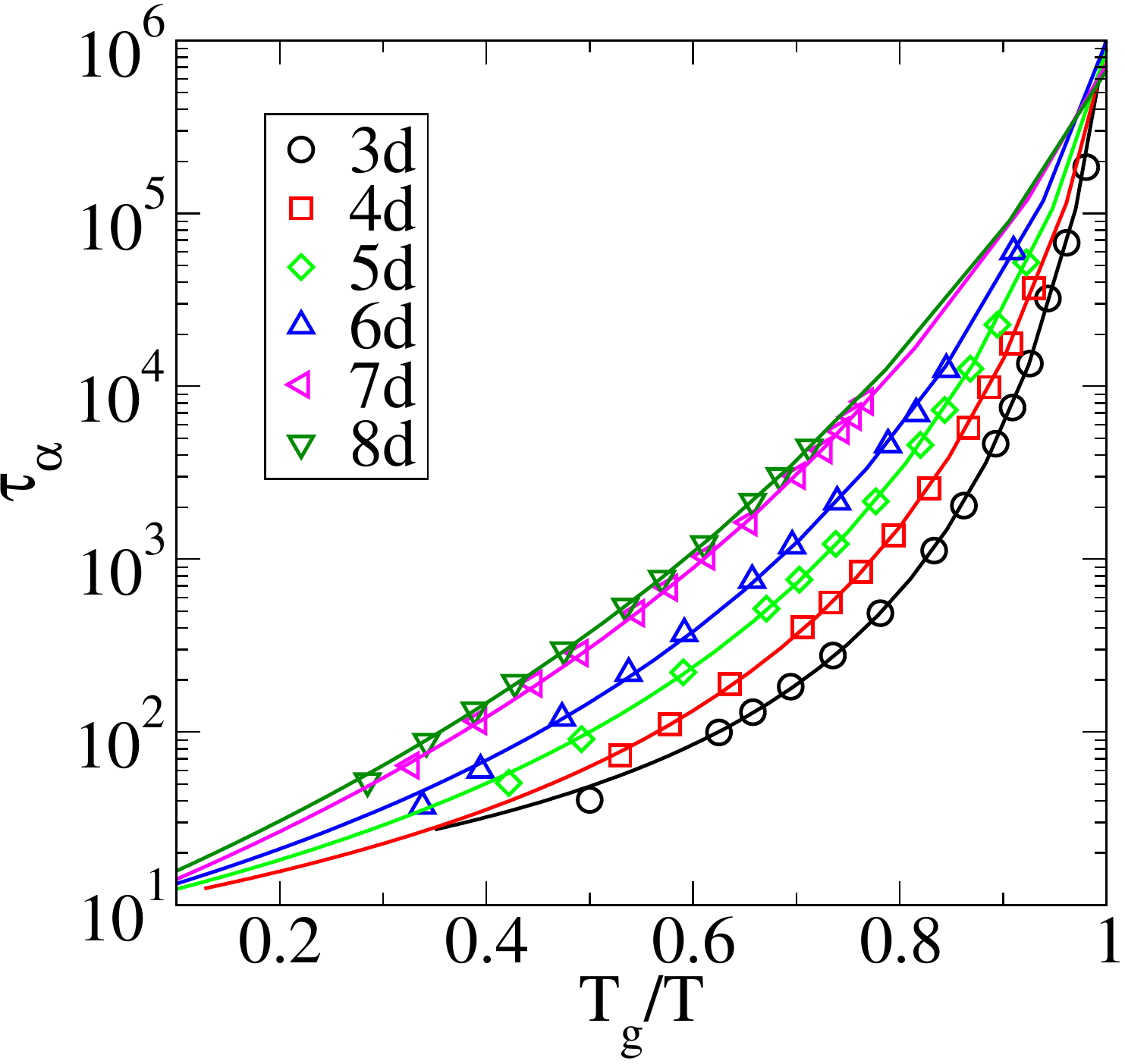}
\caption{Relaxation time as a function of temperature for various dimensions. Left: Points (dot) represent simulation data along with fits (lines) to the VFT form. Right: Angell plot, of logarithm of the relaxation times, plotted against $T/T_g$ where the glass transition temperature $T_g$ is chosen to be the temperature at which the relaxation time reaches the value $10^6$. The fragility is highest for $3d$, with the fragility decreasing with increasing spatial dimensionality.}
\label{fig-VFT}
\end{figure}

\begin{figure}[htp]
\centering
\includegraphics[width=0.41\textwidth]{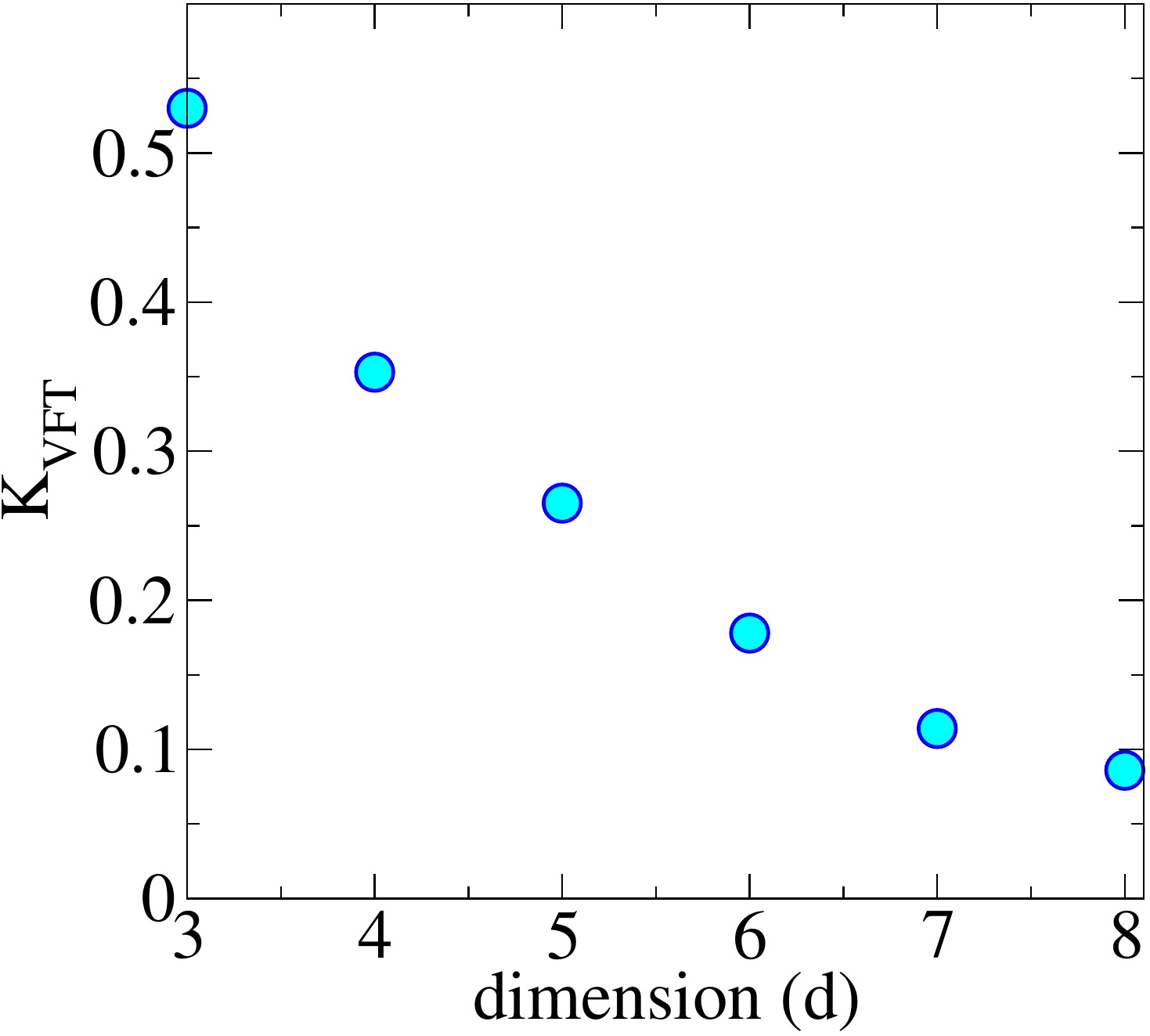}
\includegraphics[width=0.45\textwidth]{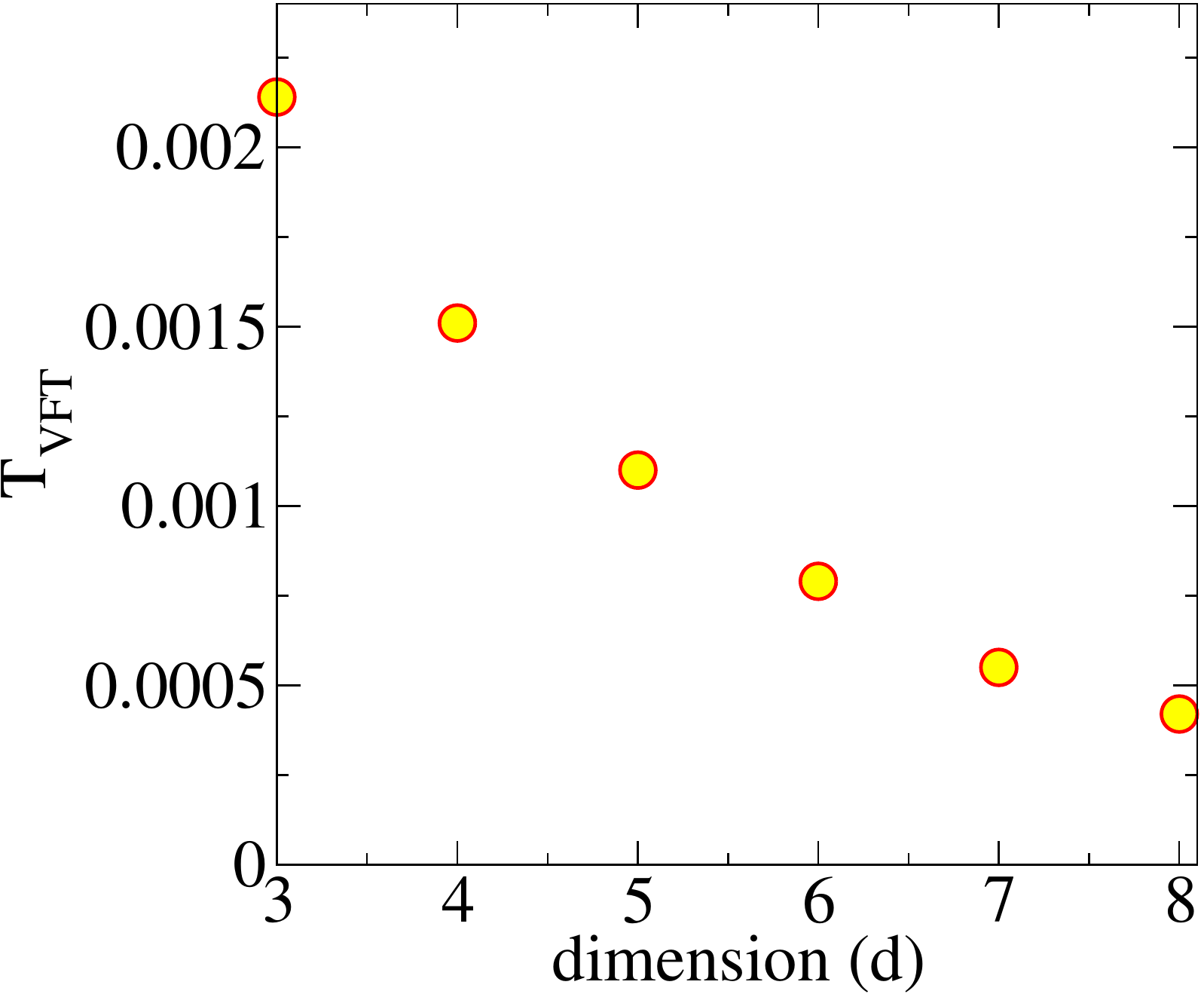}
\caption{Left: Kinetic fragility is plotted as a function of spatial dimensionality at fixed density $\phi=1.3\phi_J$, which decreases with increasing dimensionality. Right: The divergence temperature 
$T_{VFT}$ is plotted as a function of spatial dimensionality at fixed density $\phi=1.3\phi_J$ which also decreases with increasing dimensionality. }
\label{fig-frag}
\end{figure}

 We calculate the relaxation times by considering the overlap function for the $B$ particles.  The relaxation time, $\tau_{\alpha}$ is computed as the time at which $\langle q(t) \rangle =1/e$, where $\langle\cdots\rangle$ refers to an ensemble average (we average over initial times and over samples). We compute the relaxation times for a wide range of temperatures in each dimension. The relaxation times exhibit super-Arrhenius temperature dependence, the strength of which is quantified by the kinetic fragility.  
 Here, we estimate the kinetic fragility from Vogel-Fulcher-Tammann (VFT) fits to the temperature dependence of the $\alpha$ relaxation times: 
\begin{equation}
\tau_{\alpha} = \tau_0 \exp\left[\frac{1}{K_{VFT}\left(\frac{T}{T_{VFT}}-1\right)}\right],
\label{eq:vft}
\end{equation}
where $K_{VFT}$ is the kinetic fragility of the system and $T_{VFT}$ is the temperature at which the relaxation time diverges by extrapolation. In Fig. \ref{fig-VFT}, we show relaxation time as a function of temperature in a semi-log plot for each dimension, against temperature, as well as against scaled inverse temperature $T_g/T$, in an Angell plot. The {\it glass transition} temperature $T_g$ is defined as the temperature where relaxation time becomes $10^6$. The pre-factor in Eq. \ref{eq:vft}, $\tau_0 \approx 10$ in all dimensions, and thus its inclusion or otherwise in defining $T_g$ does not alter the observed behavior. From the Angell plot, it is apparent that the liquid becomes more fragile as the spatial dimension increases.  The kinetic fragility $K_{VFT}$ and the divergence temperature $T_{VFT}$ are plotted as a function of dimension in Fig. \ref{fig-frag}, which shows that both $K_{VFT}$ and $T_{VFT}$ are decreasing functions of spatial dimensionality.

\subsection{Heterogeneity in dynamics}

We next investigate the heterogeneity in dynamics by two different measures of heterogeneity: 1. The dynamical susceptibility, $\chi_4$, and  2. The non-Gaussian parameter, $\alpha_2$. 

\begin{figure*}[htp]
\centering
\includegraphics[width=0.31\textwidth]{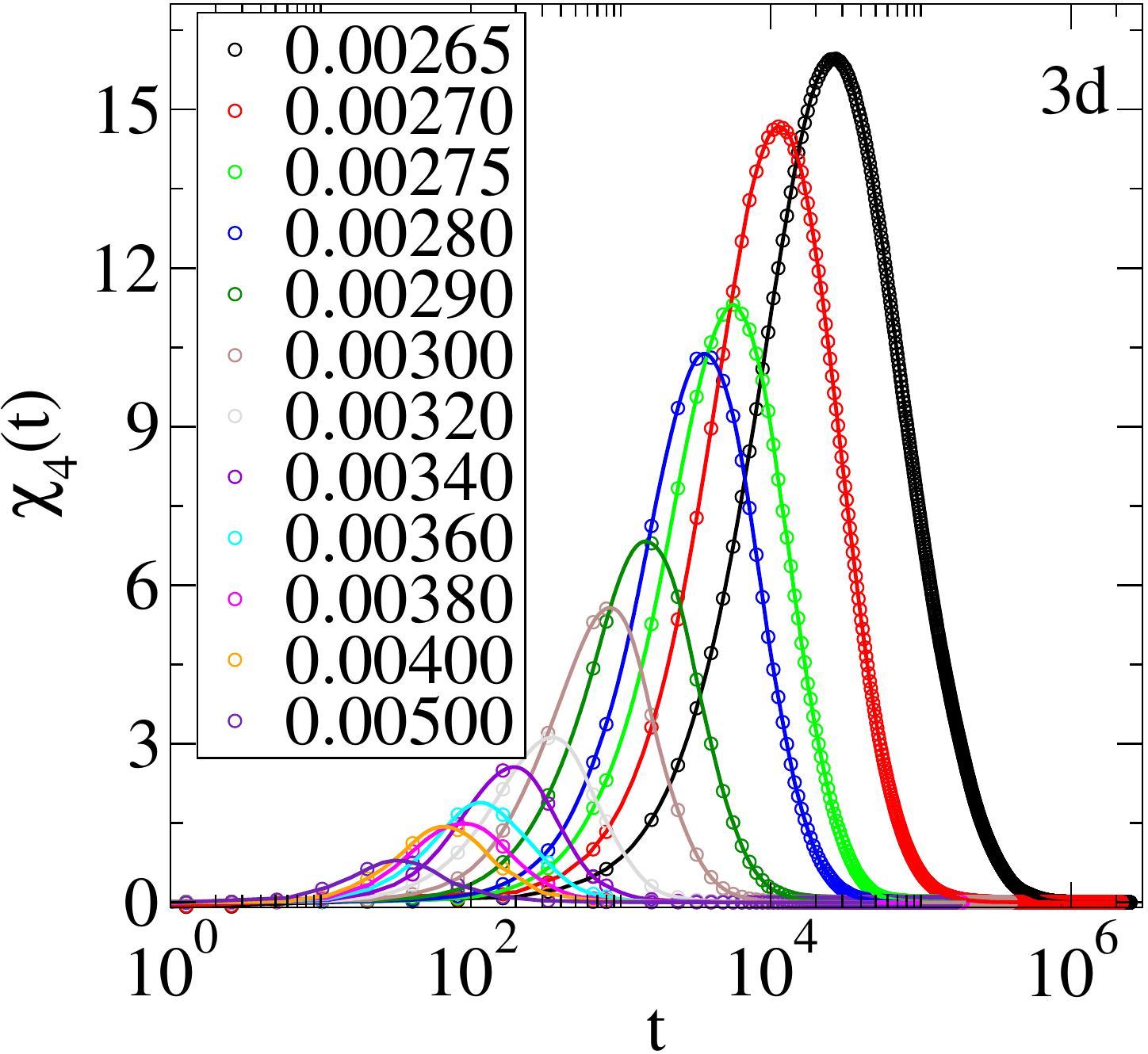}
\includegraphics[width=0.31\textwidth]{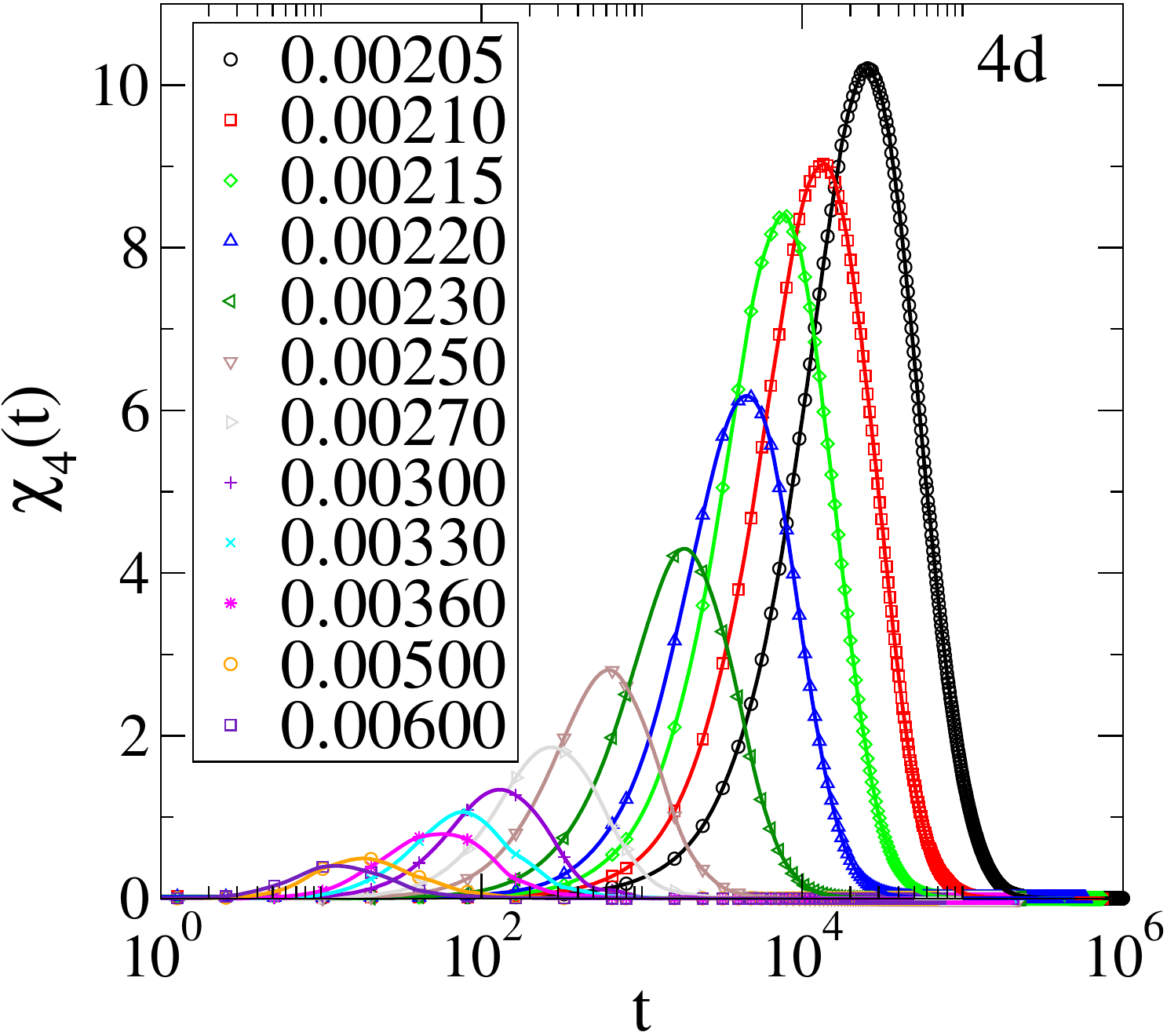}
\includegraphics[width=0.31\textwidth]{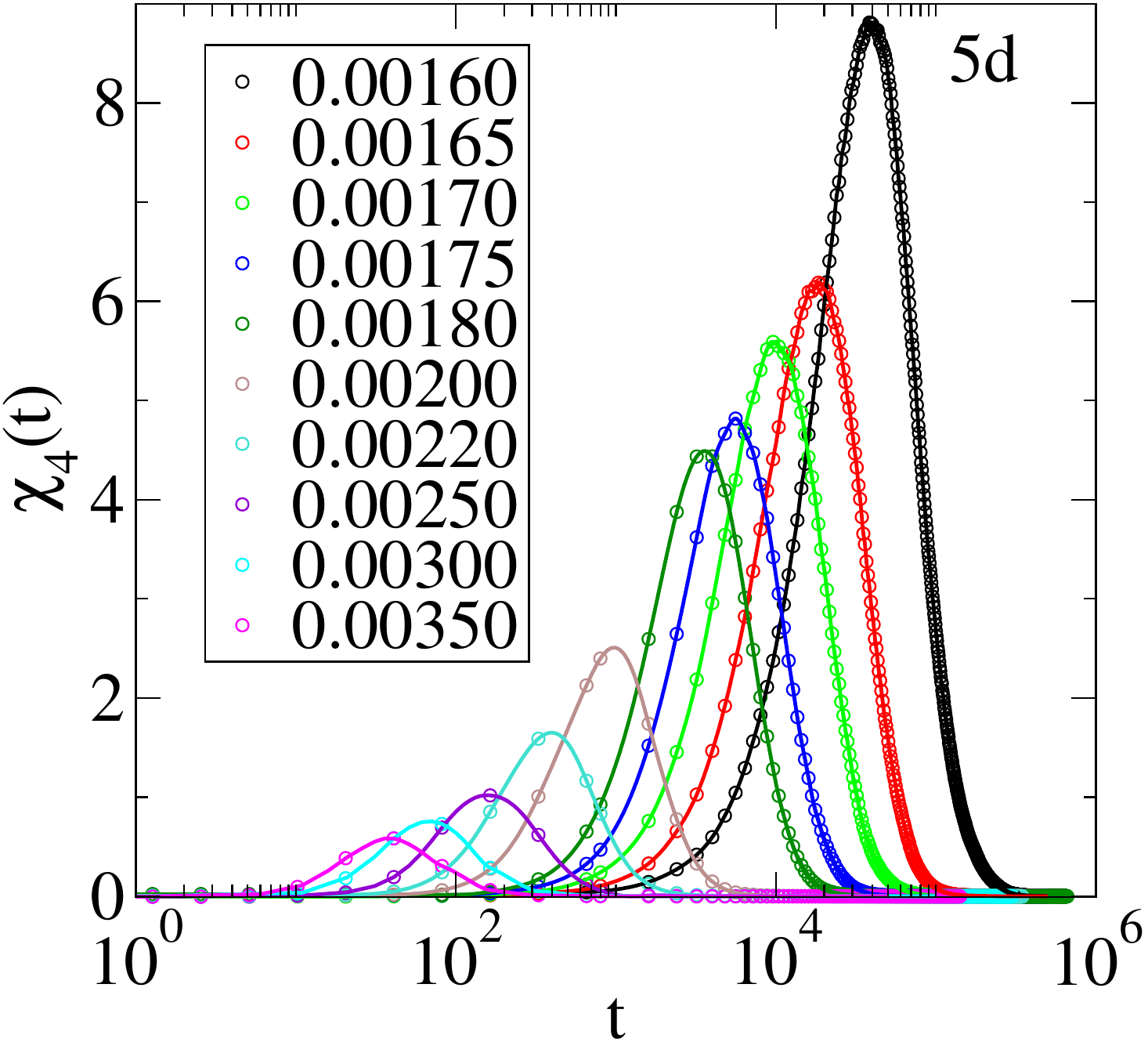}
\includegraphics[width=0.31\textwidth]{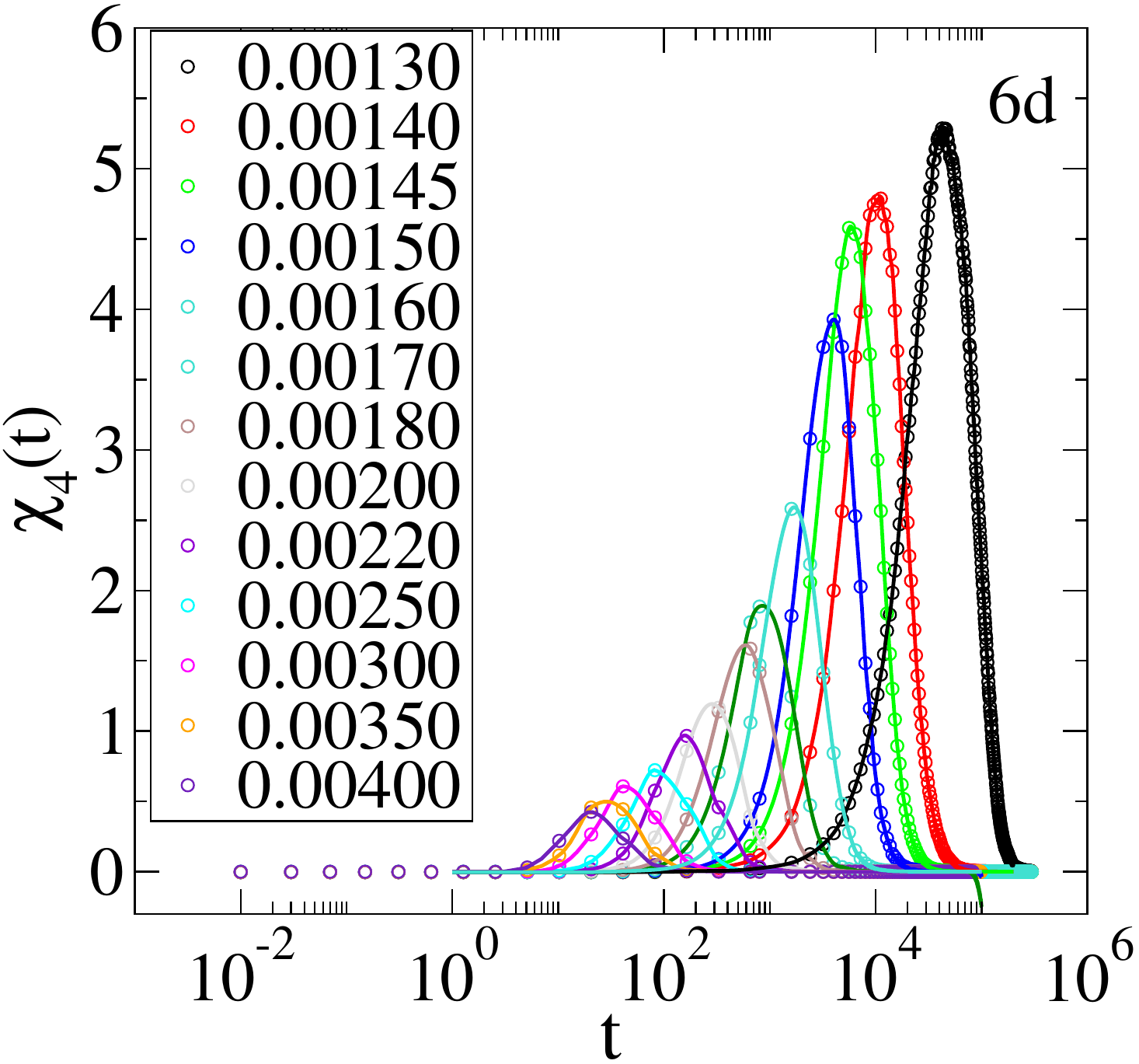}
\includegraphics[width=0.31\textwidth]{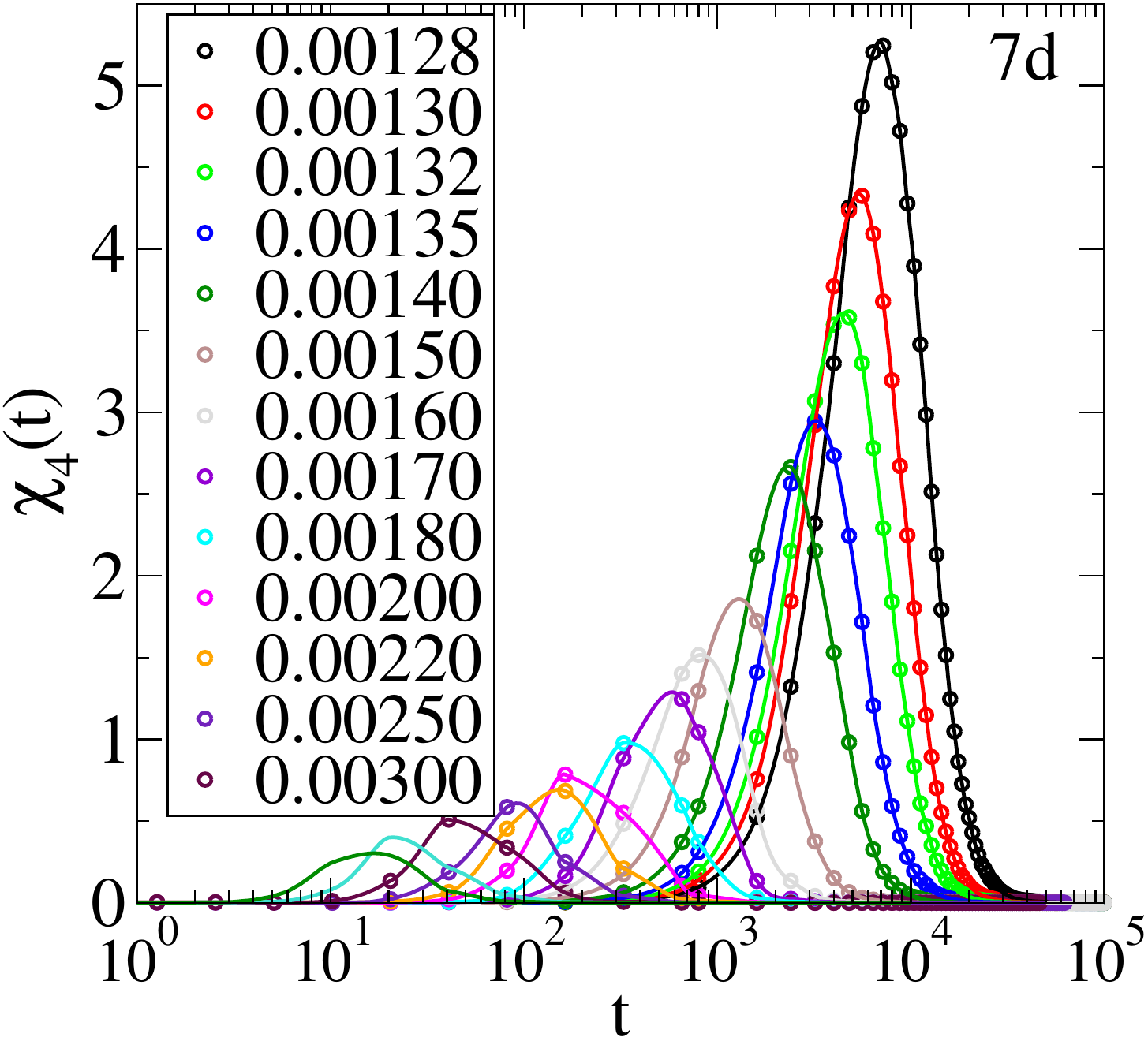}
\includegraphics[width=0.31\textwidth]{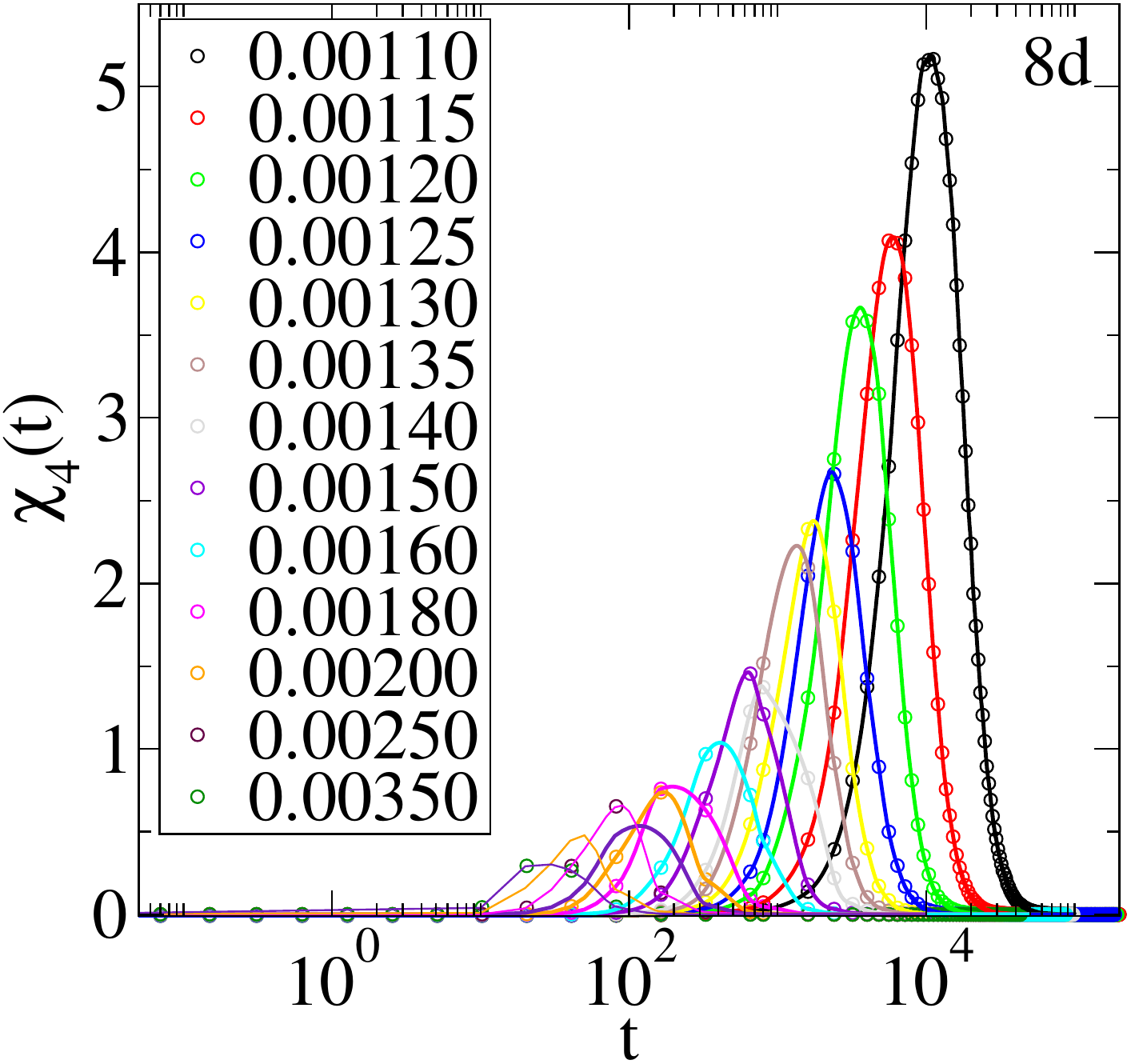}
\caption{The dynamical susceptibility $\chi_4(t)$ is plotted as a function of time for different temperatures for spatial dimensions $3-8$.}
\label{fig-chi4}
\end{figure*}

\subsubsection{Dynamical susceptibility, $\chi_4$}

Dynamical susceptibility, $\chi_4$, which measures the fluctuations in the overlap function $q(t)$, is defined by:
\begin{equation}
\chi_4(t) = N_B\left[\langle q(t)^2\rangle - \langle q(t)\rangle^2\right]
\end{equation} 

where the average is over initial configurations and the independent samples. 

\begin{figure}[htp]
\centering
\includegraphics[width=0.48\textwidth]{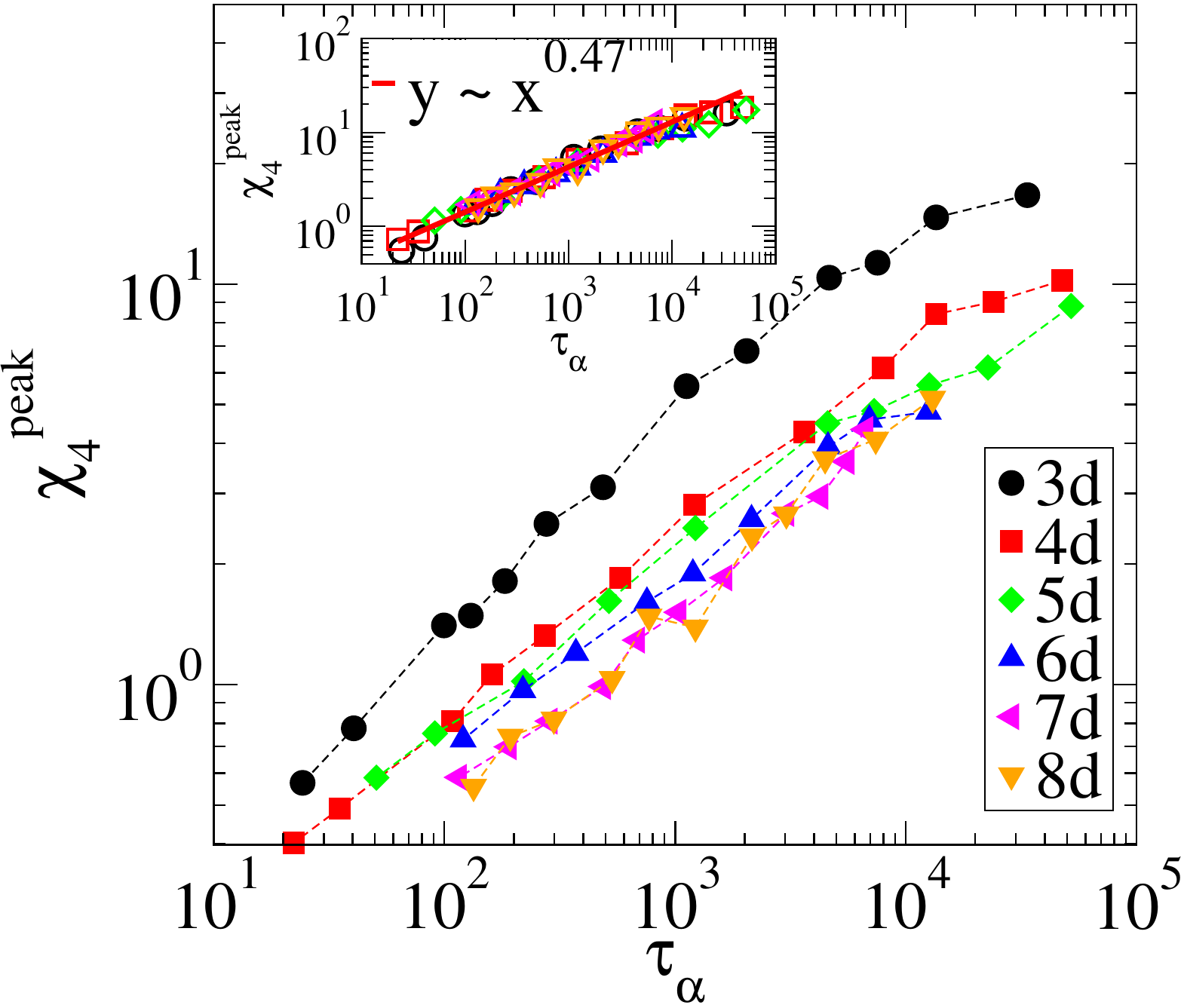}
\includegraphics[width=0.48\textwidth]{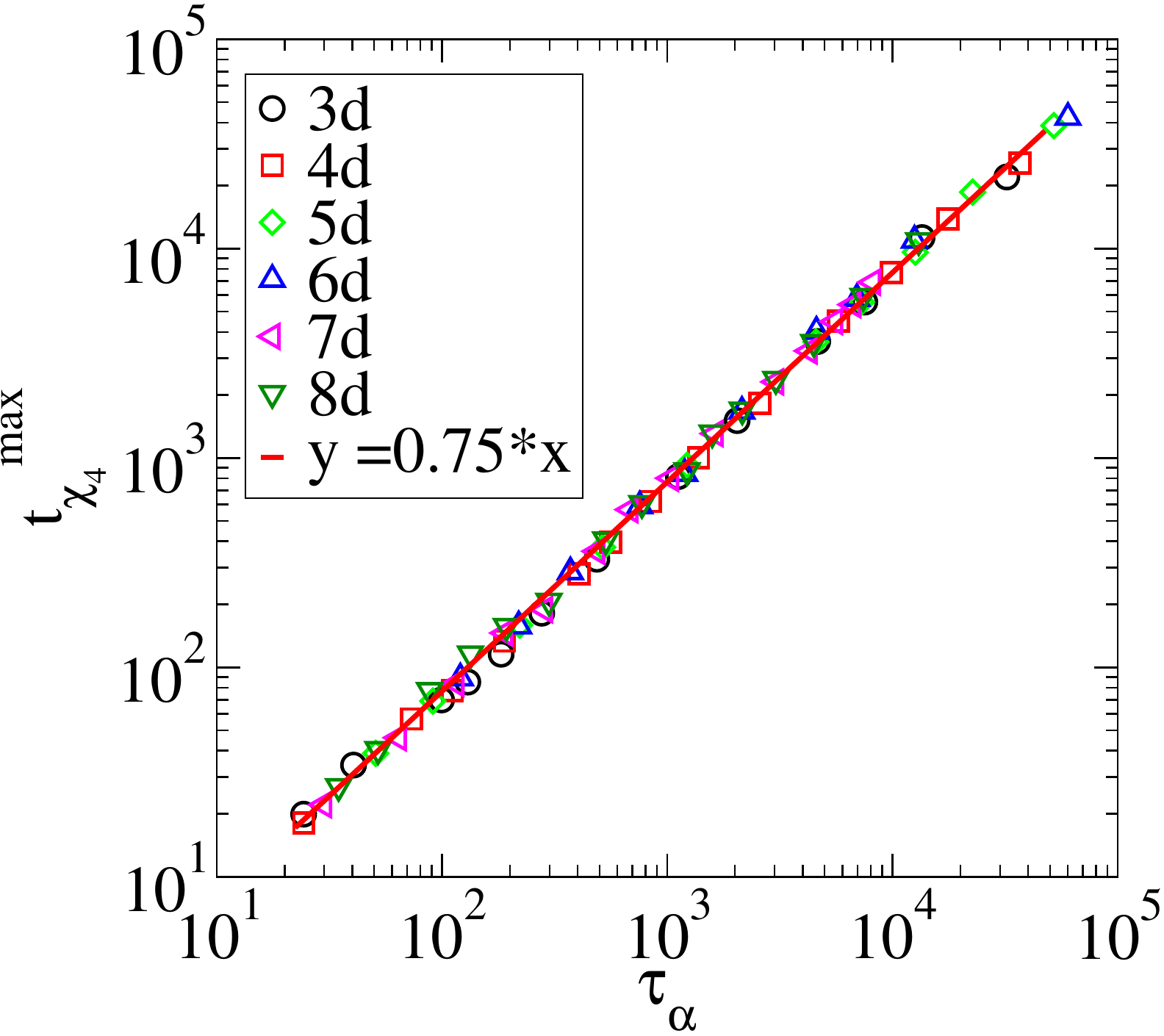}
\caption{Left: The peak value of $\chi_4(t)$, $\chi_4^{peak}$, is shown as a function of relaxation time, $\tau_{\alpha}$. Inset: By scaling $\chi_4^{peak}$ values, the data for different dimensions are collapsed onto a master curve. A power law fit (red line) provides a reasonable description for most of the temperature ($\tau_{\alpha}$) range, with exponent $0.47$. Right: The time at which $\chi_4(t)$ is maximum, $t_{\chi_4}^{max}$ is plotted as a function of $\tau_{\alpha}$ for different dimensions. The data for different dimensions overlap, and demonstrate that  $t_{\chi_4}^{max} \sim \tau_{\alpha}$.}
\label{fig-comparechi4}
\end{figure}

As it has been demonstrated extensively, the time dependence of $\chi_4(t)$ is non-monotonic, and exhibits a peak value, ($\chi_4^{peak}$), at a time that is proportional to the alpha relaxation time. 
In Fig. \ref{fig-chi4}, we show $\chi_4(t)$  against time for different temperatures in each dimension.  
The peak value of $\chi_4^{peak}$ as well as the time at which it occurs,  $t_{\chi_4}^{max}$, increase strongly upon a decrease in temperature, indicating that the heterogeneity of dynamics increases with a decrease in temperature, and is maximum at a time scale that increases in proportion to $\tau_{\alpha}$.
To compare the degree of heterogeneity for different spatial dimensions, we show, in Fig. \ref{fig-comparechi4} (Left panel), $\chi_4^{peak}$ as a function of $\tau_{\alpha}$ for each dimension. For a given $\tau_{\alpha}$,  $\chi_4^{peak}$  decreases with increasing spatial dimension implying that  heterogeneity decreases with increasing spatial dimensionality. We also observe that $\chi_4^{peak}$ shows a power law dependence on $\tau_{\alpha}$ at higher temperatures as $\chi_4^{peak} \sim \tau_{\alpha}^z$, with $z$ being the power-law exponent. Deviations from power law behaviour is mostly observed at lower temperatures. This behaviour is consistent with previous observations in three dimensions ~\cite{karmakar2009growing,flenner2010dynamic, flenner2014universal,lavcevic2003spatially}, including polymeric glass formers \cite{xu2020molecular} and exponent $z$ is found to be close to $0.62$ for binary hard sphere fluids \cite{flenner2010dynamic, flenner2014universal} whereas $z = 0.51$ has recently been found for a model liquid that aims to tune the degree of mean field character \cite{nandi2021connecting}. Our estimate is on the lower side of these values, but close to that reported in \cite{nandi2021connecting}. Since the power law regime is limited in extent, there is room for error in the exact determination of the exponent. Remarkably, however, we find that the exponent of the power law, $z$, is the same in all dimensions for our studied model, as evidenced by the data collapse, regardless of the precise value. The expectation of a power law dependence arises, for example, from inhomogeneous mode coupling theory \cite{biroli2006inhomogeneous}, and the deviations from the power law are understood to be a consequence of the role played by activated processes at low temperatures. Thus, the observation of a common exponent describing the power law dependence of  $\chi_4^{peak}$ on $\tau_{\alpha}$ should perhaps be seen as mean-field behaviour that does not depend on spatial dimensions and thus not surprising. Nevertheless, to our knowledge, such a universal behaviour has not previously been reported across the range of spatial dimensions that we investigate.

\subsubsection{Non-Gaussian parameter, $\alpha_2(t)$}

\begin{figure*}[htp]
\centering
\includegraphics[width=0.31\textwidth]{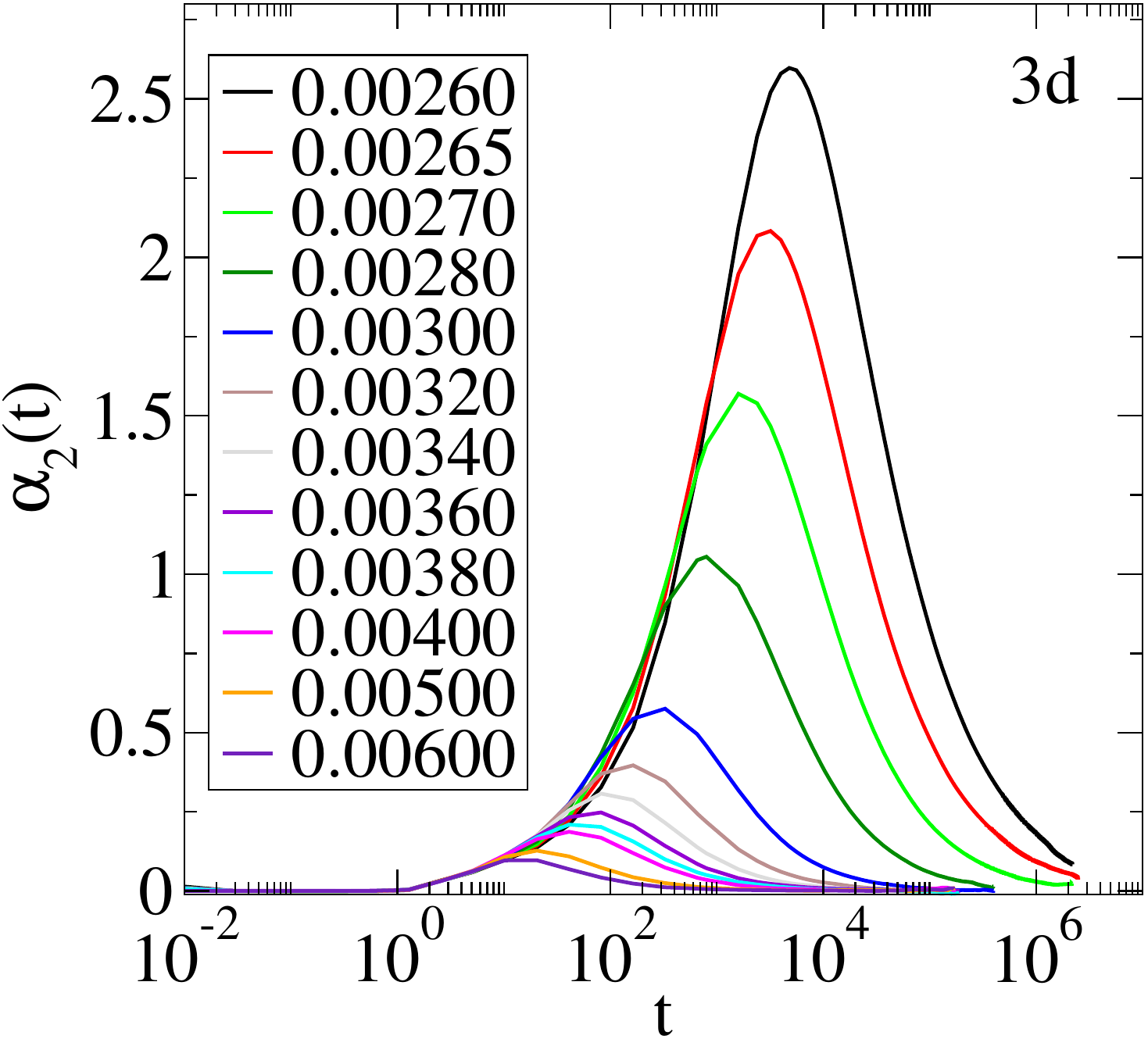}
\includegraphics[width=0.31\textwidth]{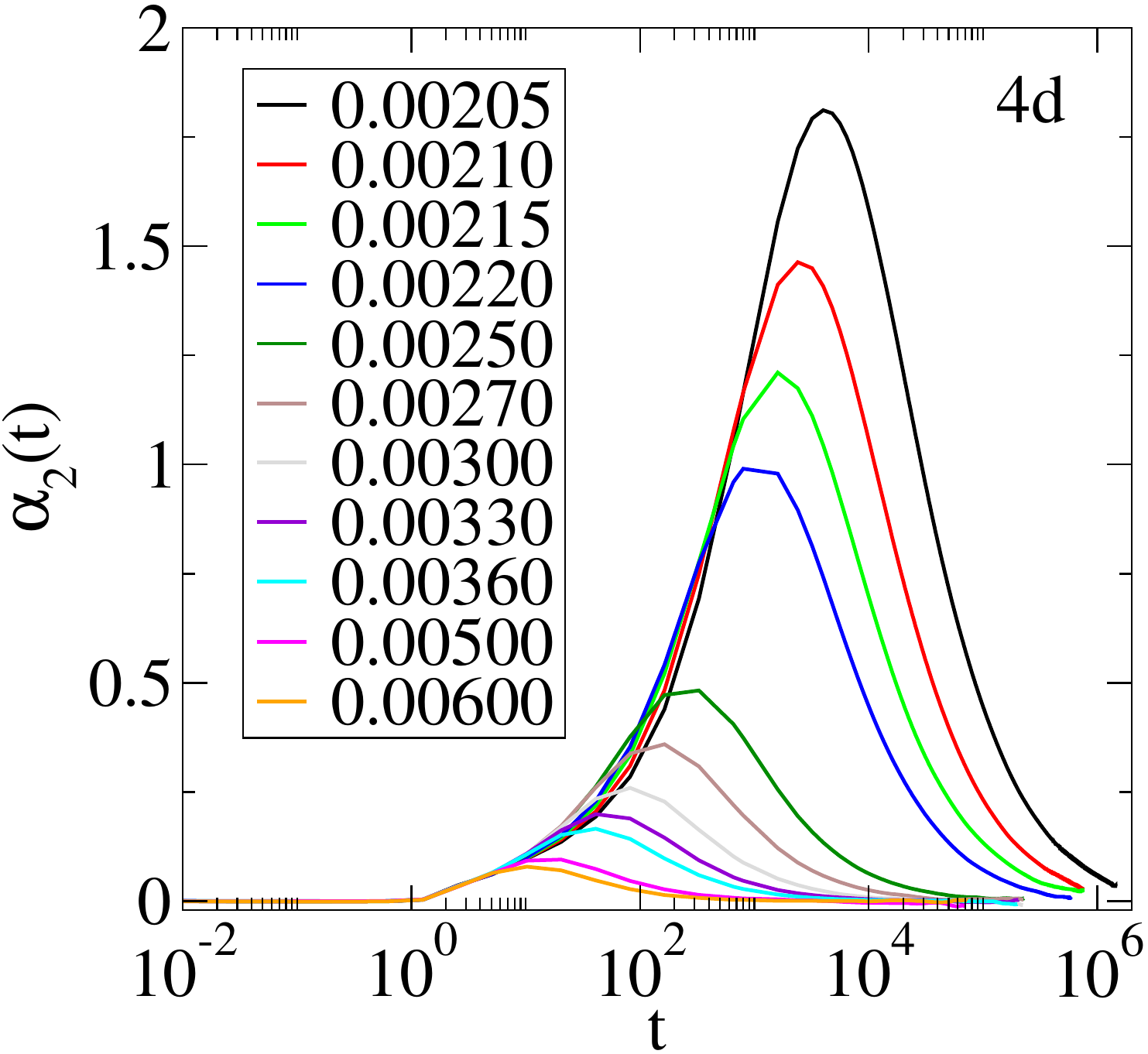}
\includegraphics[width=0.31\textwidth]{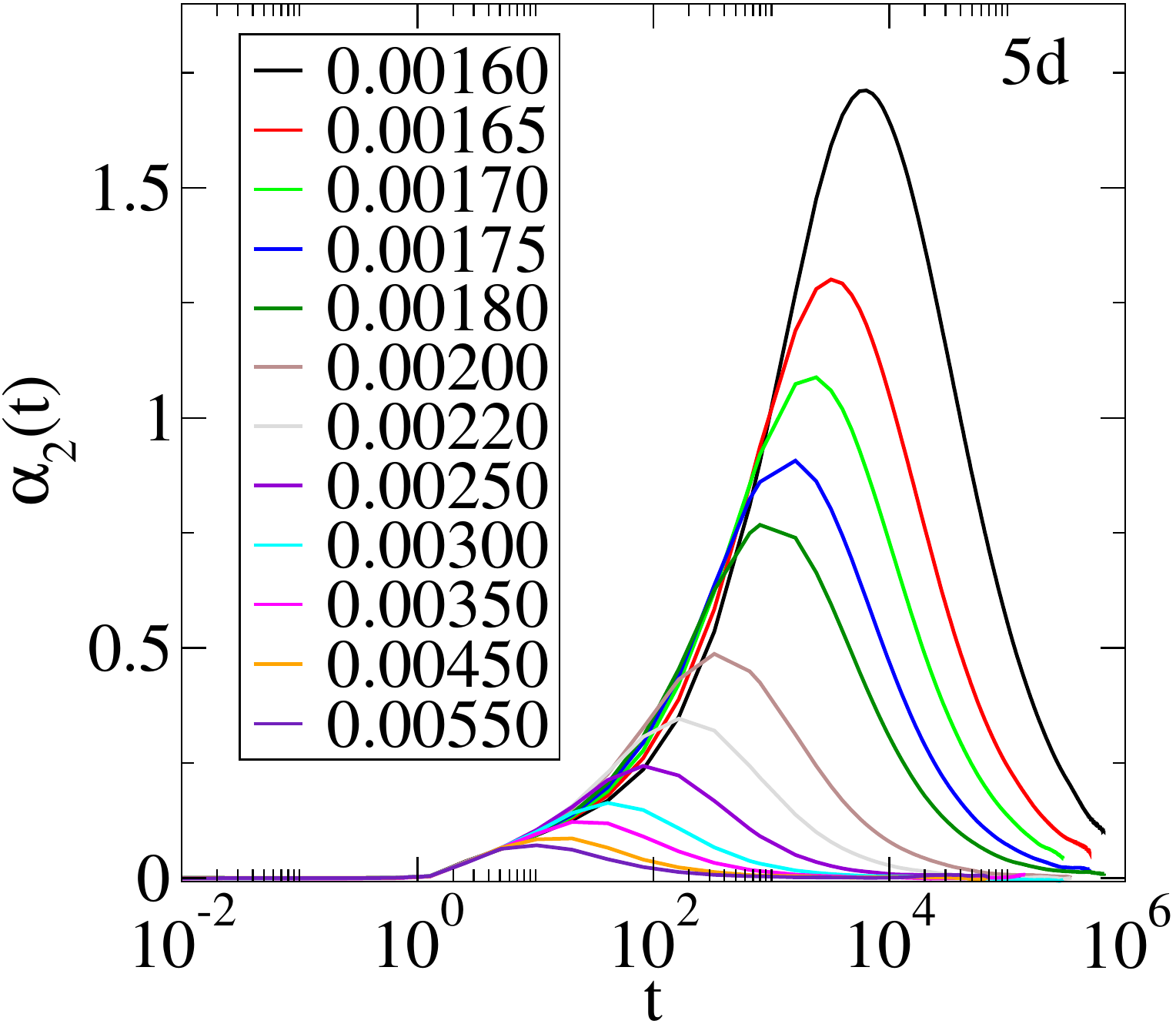}
\includegraphics[width=0.31\textwidth]{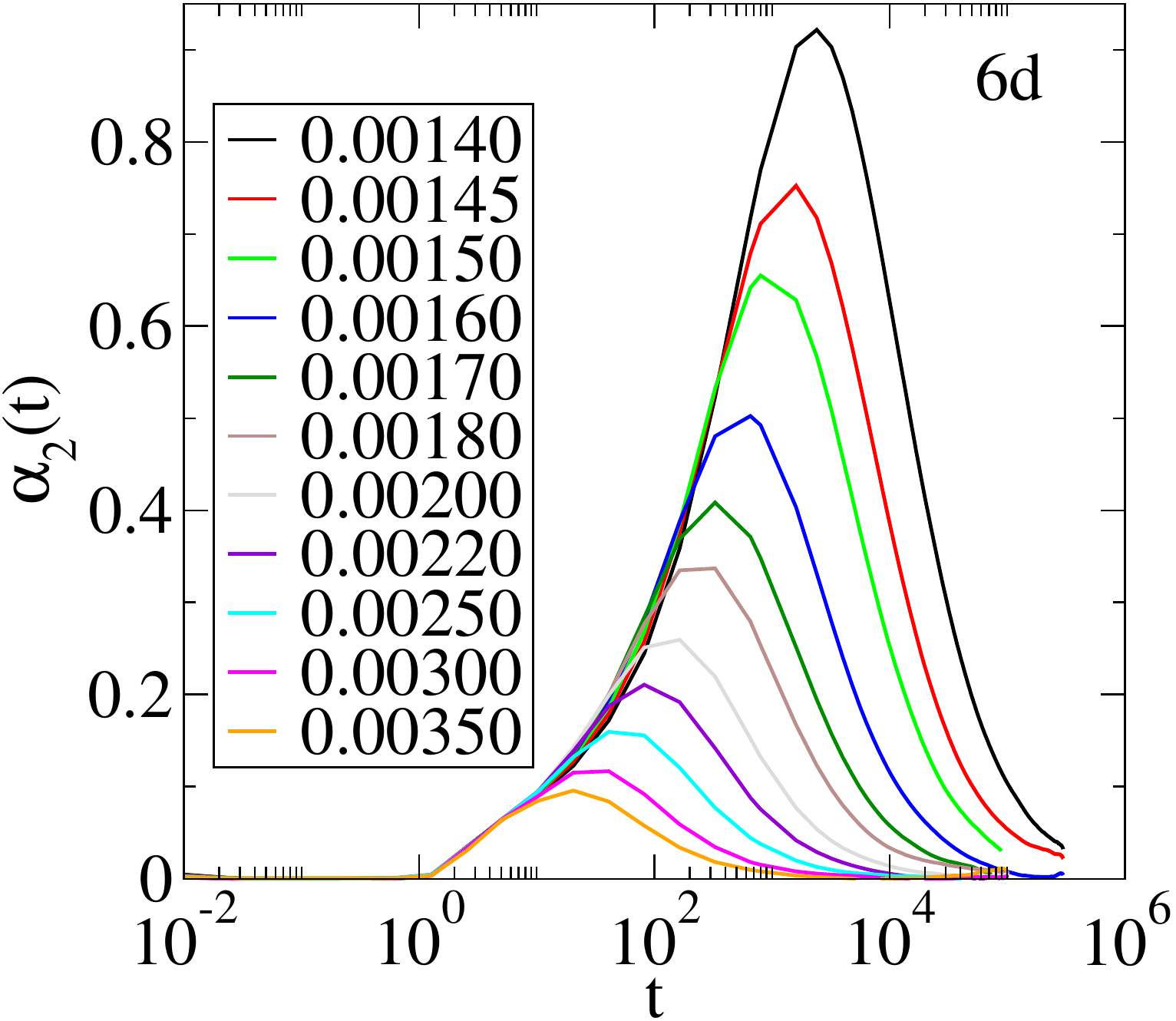}
\includegraphics[width=0.31\textwidth]{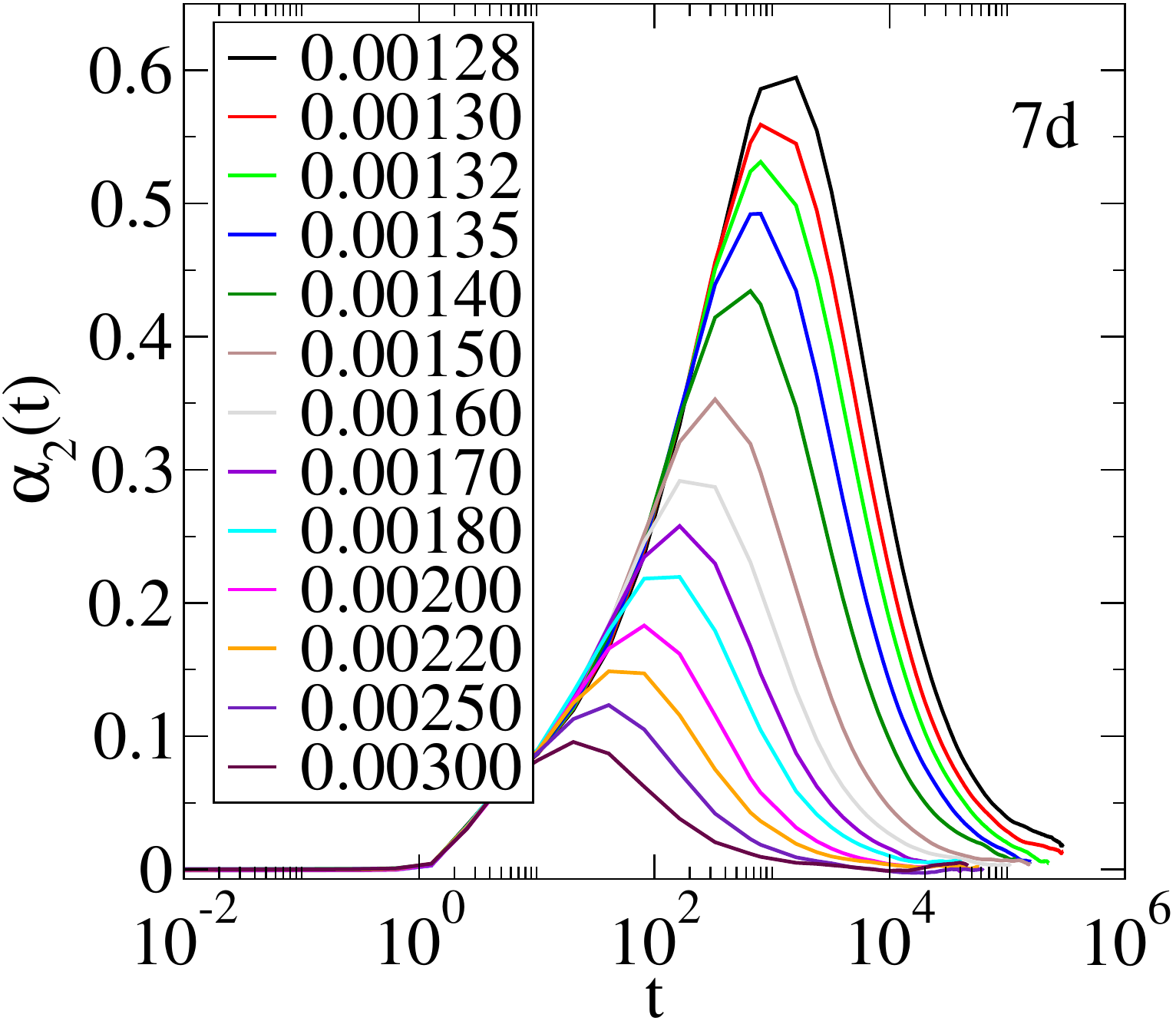}
\includegraphics[width=0.31\textwidth]{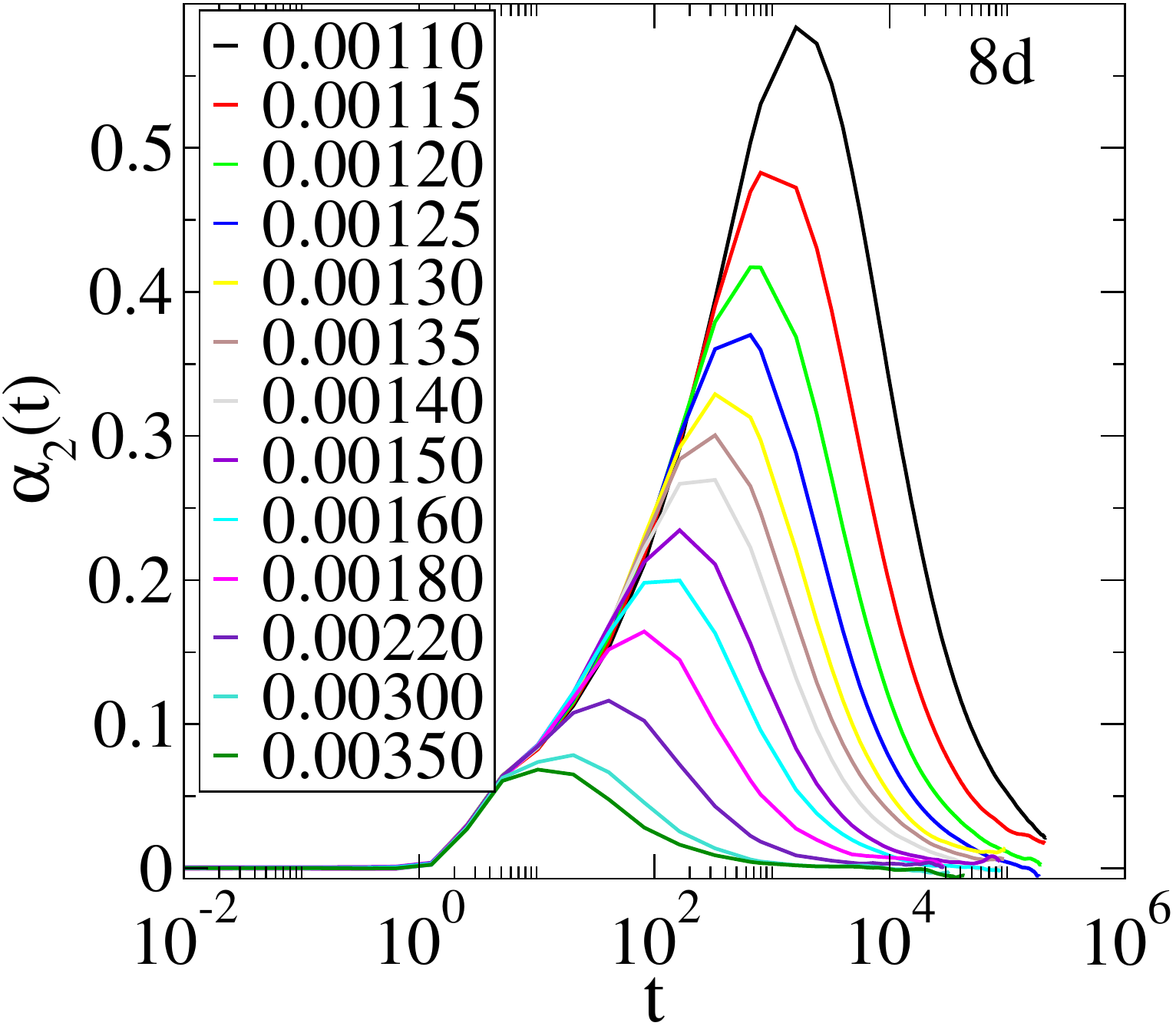}
\caption{The non-Gaussian parameter $\alpha_2$ is plotted against time for different temperatures and  spatial dimensions $3-8$.}
\label{fig-alpha2}
\end{figure*}

\begin{figure}[htp]
\centering
\includegraphics[width=0.46\textwidth]{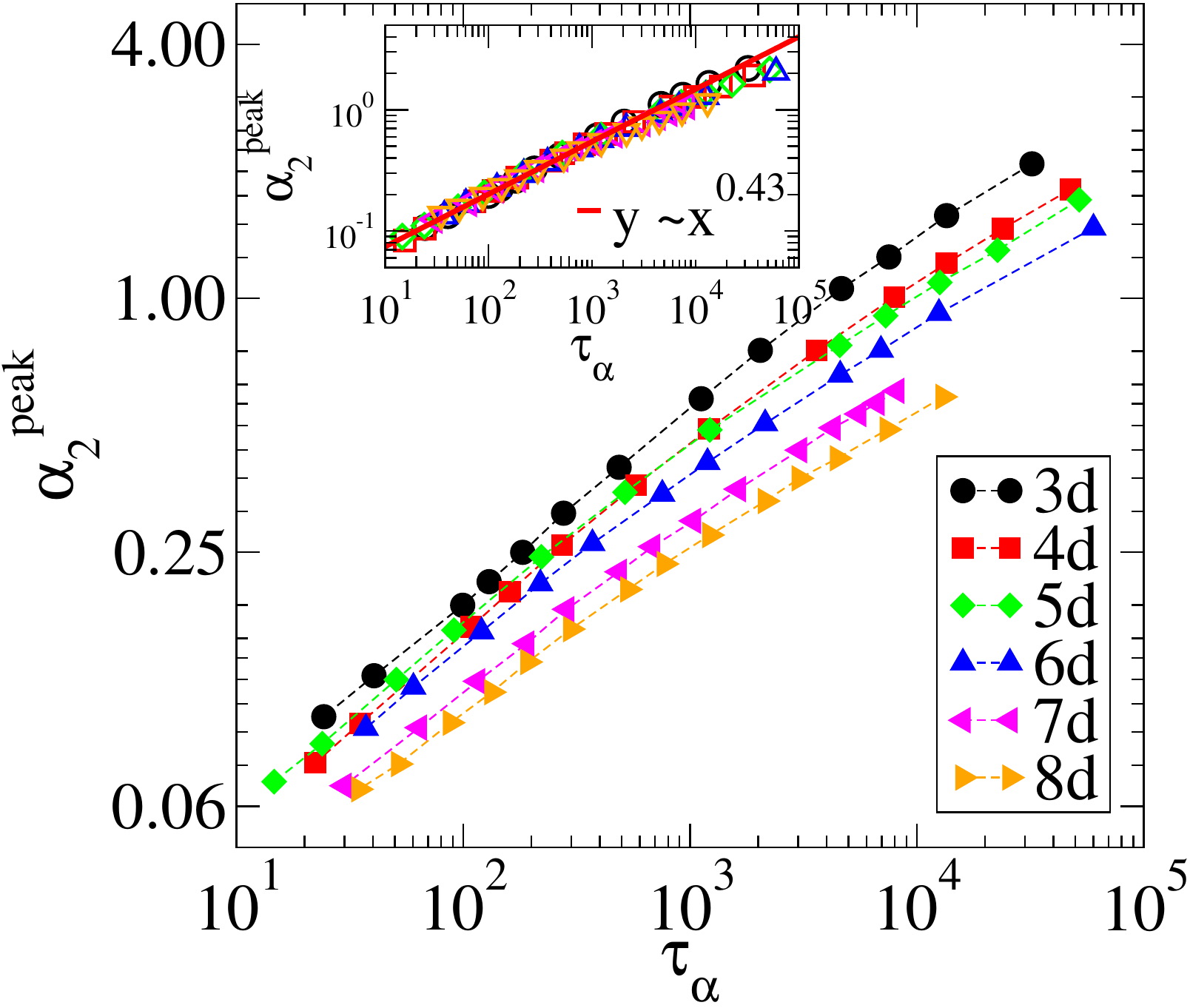}
\includegraphics[width=0.44\textwidth]{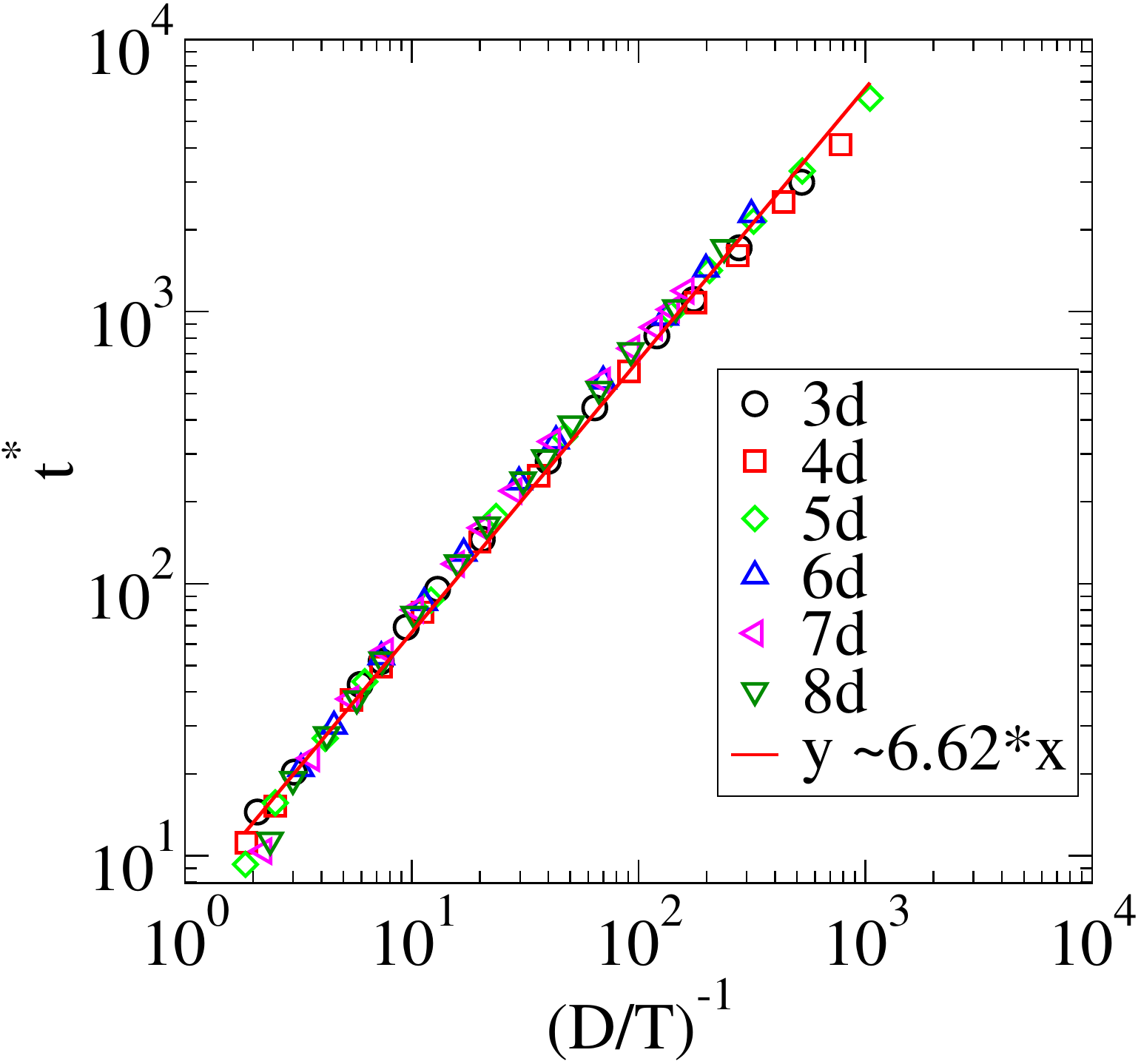}
\caption{Left: The peak value of $\alpha_2(t)$, $\alpha_2^{peak}$, is shown as a function of relaxation time, $\tau_{\alpha}$ for different dimensions. Inset: By scaling $\alpha_2^{peak}$ values,  
the data for different dimensions are collapsed onto a master curve. A power law fit (red line) provides a reasonable description for most of the temperature ($\tau_{\alpha}$) range, with exponent $0.43$. 
Right: The time at which $\alpha_2(t)$ is maximum, $t^{*}$ is plotted as a function of $(D/T)^{-1}$
for different dimensions. The data for different dimensions overlap, and demonstrate that  
$t^* \sim (D/T)^{-1}$, albeit with small deviations apparent at low temperatures.}
\label{alpha2-compare}
\end{figure}

Next, we investigate another measure of heterogeneity, the non-Gaussian parameter, $\alpha_2(t)$, for different dimensions. As previously discussed in detail \cite{starr2013relationship,xu2016influence,xu2020molecular}, $\alpha_2(t)$ and $\chi_4$ correspond to distinct aspects of heterogeneity, associated with correlated clusters of mobile, and immobile, particles respectively.. The non-Gaussian parameter, $\alpha_2(t)$ measures the deviation of the van Hove distribution of displacements of the particle in time $t$ from Gaussian form, expected for spatially homogeneous dynamics, and is given by 

\begin{eqnarray}
\alpha_2(t) &=& C_d \frac{\langle r^4(t)\rangle}{\langle r^2(t) \rangle^2} -1 \\
\langle r^{2n}\rangle &=&\frac{1}{N} \langle (\Vec{r}_i(t)- \vec{r}_i(0))^{2n}\rangle
\end{eqnarray}
where $C_d$ is a spatial dimension dependent coefficient to ensure that $\alpha_2(t) = 0$ when the distribution of displacements is a Gaussian. Similar to $\chi_4$, $\alpha_2$ also shows non-monotonic behaviour with respect to time. However, the characteristic time $t^{*}$ at which $\alpha_2(t)$ is maximum is smaller than $\tau_{\alpha}$, and has been demonstrated to be proportional to a time scale determined by the diffusion coefficient, $(D/T)^{-1}$ \cite{starr2013relationship}.
In Fig.  \ref{fig-alpha2}, we show $\alpha_2(t)$ against time for different temperatures and spatial dimensions $3-8$. We see that $\alpha_2^{peak}$ increases with a decrease in temperature for all spatial dimensions. Similarly to $\chi_4$, we report $\alpha_2^{peak}$ against $\tau_{\alpha}$ for spatial dimension $3-8$ in Fig. \ref{alpha2-compare} (Left panel).  We see that for a given $\tau_{\alpha}$, $\alpha_2^{peak}$ also decreases with increasing dimensionality. This again implies that heterogeneity decreases with increasing spatial dimensionality. Similar to $\chi_4^{peak}$, $\alpha_2^{peak}$ also displays a  power law dependence on $\tau_{\alpha}$ at higher dimensions, with deviations at lower temperatures. The inset of Fig. \ref{alpha2-compare} (Left panel) demonstrates that for $\alpha_2^{peak}$, the exponent of the power law is  $0.43$ and is a good description of the data for all dimensions. In \cite{wang2018revealing} the behaviour of $\alpha_2^{peak}$ {\it vs.} $\tau_{\alpha}$ was fitted to two power laws, with exponents $0.8$ (high temperatures) and $0.3$ (low temperatures). While we find the exponent to be $0.43$ convincingly over two decades of (high to moderate temperature) relaxation times, we do find that an exponent of $0.3$ is a good description of low temperature data. In Fig. \ref{alpha2-compare} (Right panel),  we show the time $t^{*}$ against $(D/T)^{-1}$, where $D$ is diffusivity, for different spatial dimensions, confirming the validity of the relation $t^{*} \sim (D/T)^{-1}$ beyond three dimensions ~\cite{starr2013relationship}. The observed relationship between $t^*$ and $D/T$ has been found to be valid in many different glass formers as well as other materials \cite{starr2013relationship,sengupta2013breakdown, wang2019universal, zhang2019superionic}, and our results show that it is valid in different dimensions as well. We note, however, that an exponent other than $-1$ has been reported recently for a metallic glass former \cite{zhang2021dynamic}.

\subsection{The Breakdown of the Stokes-Einstein Relation}

A much studied phenomenon associated with glassy behavoiur is the violation or breakdown of the  Stokes-Einstein relation (SER), which relates the translational diffusion coefficient ($D$) of a Brownian particle to the shear viscosity $\eta$ of the surrounding liquid at a temperature T: $D=mk_BT/c\pi R \eta$, where $m$ is the mass and $R$ is the radius of the particle, $T$ is the temperature of the liquid, and the factor $c$ is a constant which depends on the boundary condition at the surface of the Brownian particle. It is been observed in several investigations that the SER is also satisfied when one considers the self-diffusion of particles in a liquid at relatively high temperatures (The caveats and the extent to which such a statement is valid have also been discussed, {\it e. g.} \cite{charbonneau2013dimensional}). However, as temperature is decreased towards the glass transition, the SER is observed to break down. 
\marktext{As mentioned in the introduction, violations of the SER, which can be expressed as ${D \eta \over T} = constant$, have been investigated considering $\tau_{\alpha}$ in place of ${\eta \over T}$, expressing the SER as $D \tau_{\alpha} = constant$ \cite{sengupta2013breakdown,charbonneau2013dimensional,parmar2017length}. This equivalence has been validated by computing the viscosity $\eta$ and comparing with $\tau_{\alpha}$, at the wave vector corresponding to the peak of the structure factor \cite{sengupta2013breakdown}. Further, several works have considered relaxation behaviour as a function of the wave vector $k$  \cite{kim2005breakdown,sengupta2013breakdown,charbonneau2013dimensional,parmar2017length}, either through the $k$ dependent viscosity \cite{kim2005breakdown} or relaxation times computed as a function of $k$. In \cite{parmar2017length}, it was shown that for a given $k$, violation of the SER arise when $k^{-1}$ falls below a length scale characterising dynamical heterogeneity. In the present work, we do not investigate the $k$ dependence of the violation of SER, but consider only the $\tau_{\alpha}$ defined above,  which empirically is equivalent to considering the SER between diffusion coefficients and viscosity, as mentioned above. 
The enhancement of the diffusion coefficient was obtained  by Kim and Keyes \cite{kim2005breakdown} through a mode coupling expression for the diffusion coefficient as an integral over the inverse of the $k$ dependent viscosity $\eta(k)$. Thus, the reduction of $\eta(k)$ with respect to the hydrodynamic value is offered as a compelling explanation of the violation of the SER. Indeed, intuitively,  the observations employing relaxation times obtained as a function of wave-vector  \cite{sengupta2013breakdown,charbonneau2013dimensional,parmar2017length} are consistent with such an explanation. An investigation of the relationship between these different approaches has not, however, been systematically carried out, and would be interesting to perform.  The breakdown of SER is characterized by an exponent $\omega$ that describes a scaling $D \sim \tau_{\alpha}^{-1+\omega}$. As mentioned above, a limited number of previous studies  \cite{eaves2009spatial,sengupta2013breakdown,charbonneau2013dimensional} have considered SER and the breakdown thereof as a function of spatial dimension.   Charbonneau \emph{et al.} \cite{charbonneau2013dimensional} have performed a hydrodynamic analysis of SER for varying spatial dimension, as well as numerical investigations up to $d = 10$ for hard sphere liquids. Here, we examine validity or breakdown of the SER employing the self diffusion coefficients $D$ and the $\tau_{\alpha}$ described above, as a function of spatial dimensionality.}

\begin{figure}[htp]
\centering
\includegraphics[width=0.30\textwidth]{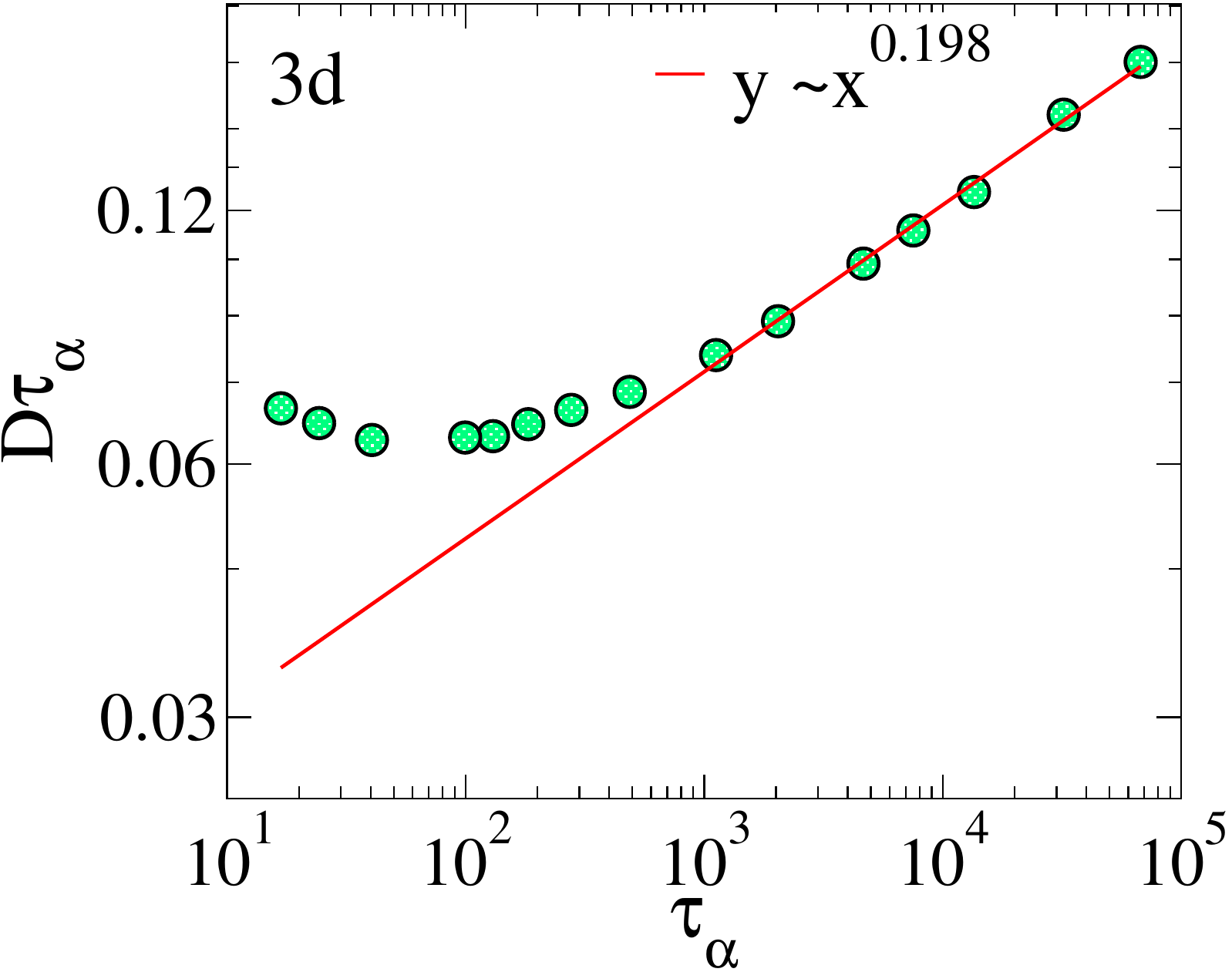}
\includegraphics[width=0.30\textwidth]{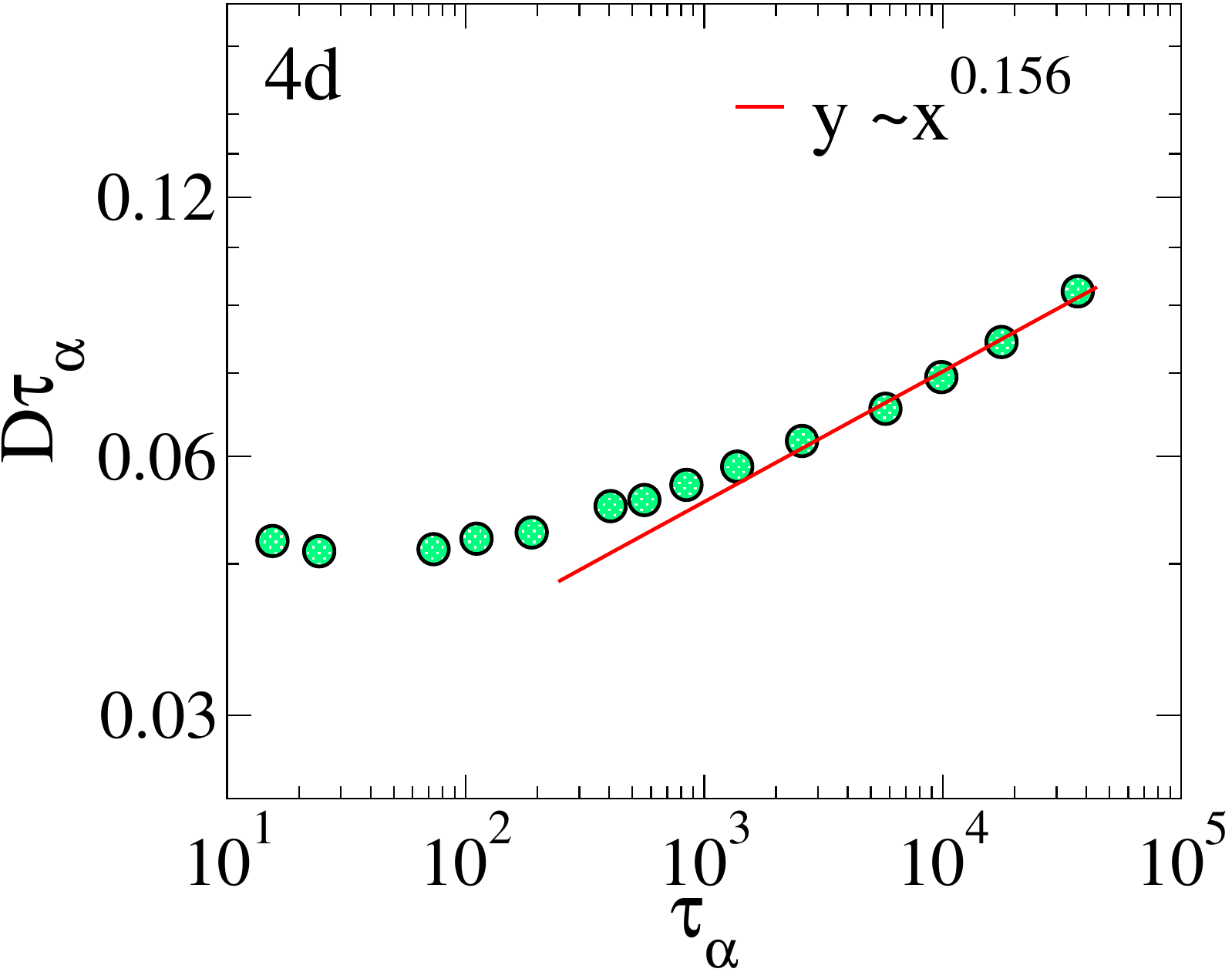}
\includegraphics[width=0.30\textwidth]{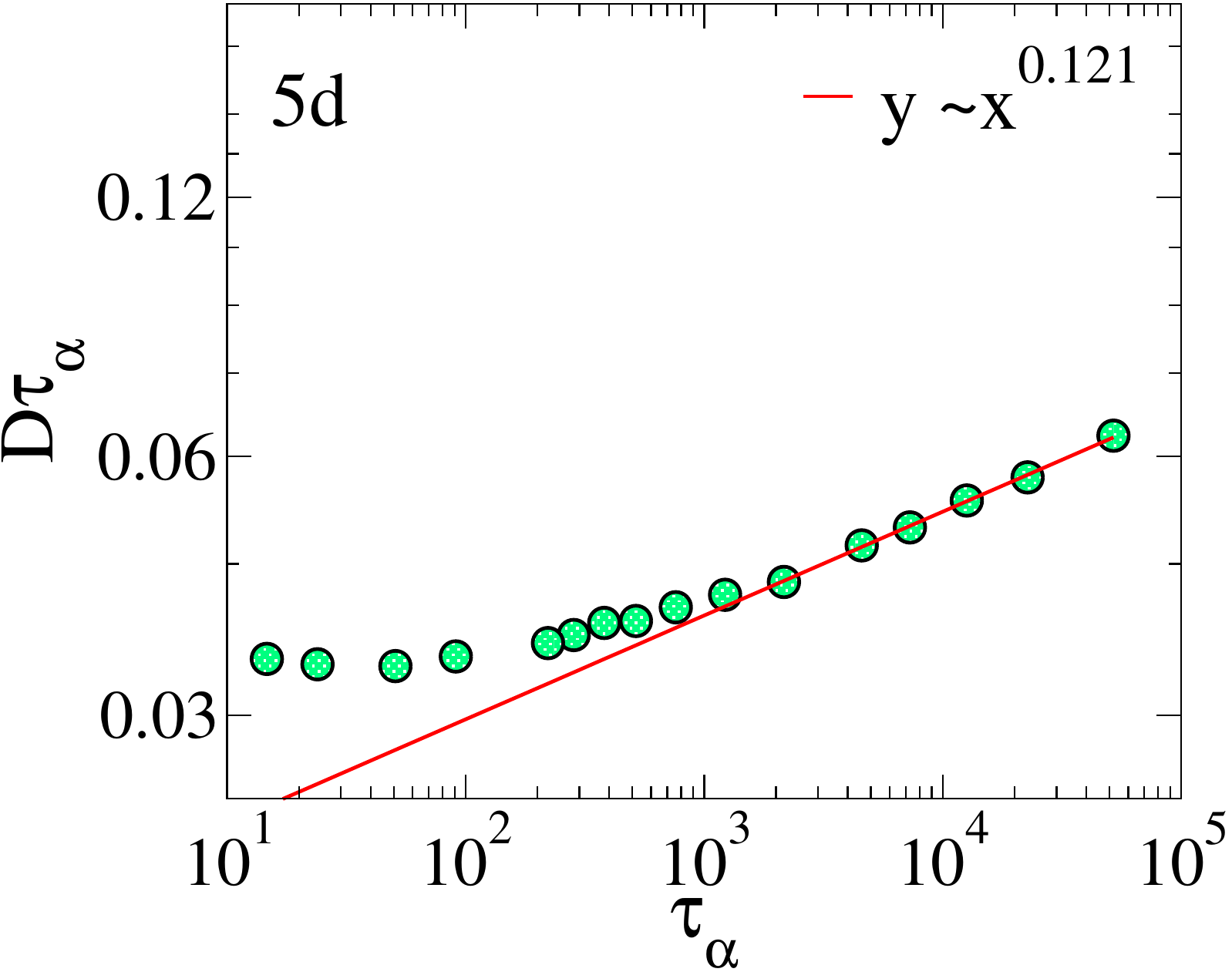}
\includegraphics[width=0.30\textwidth]{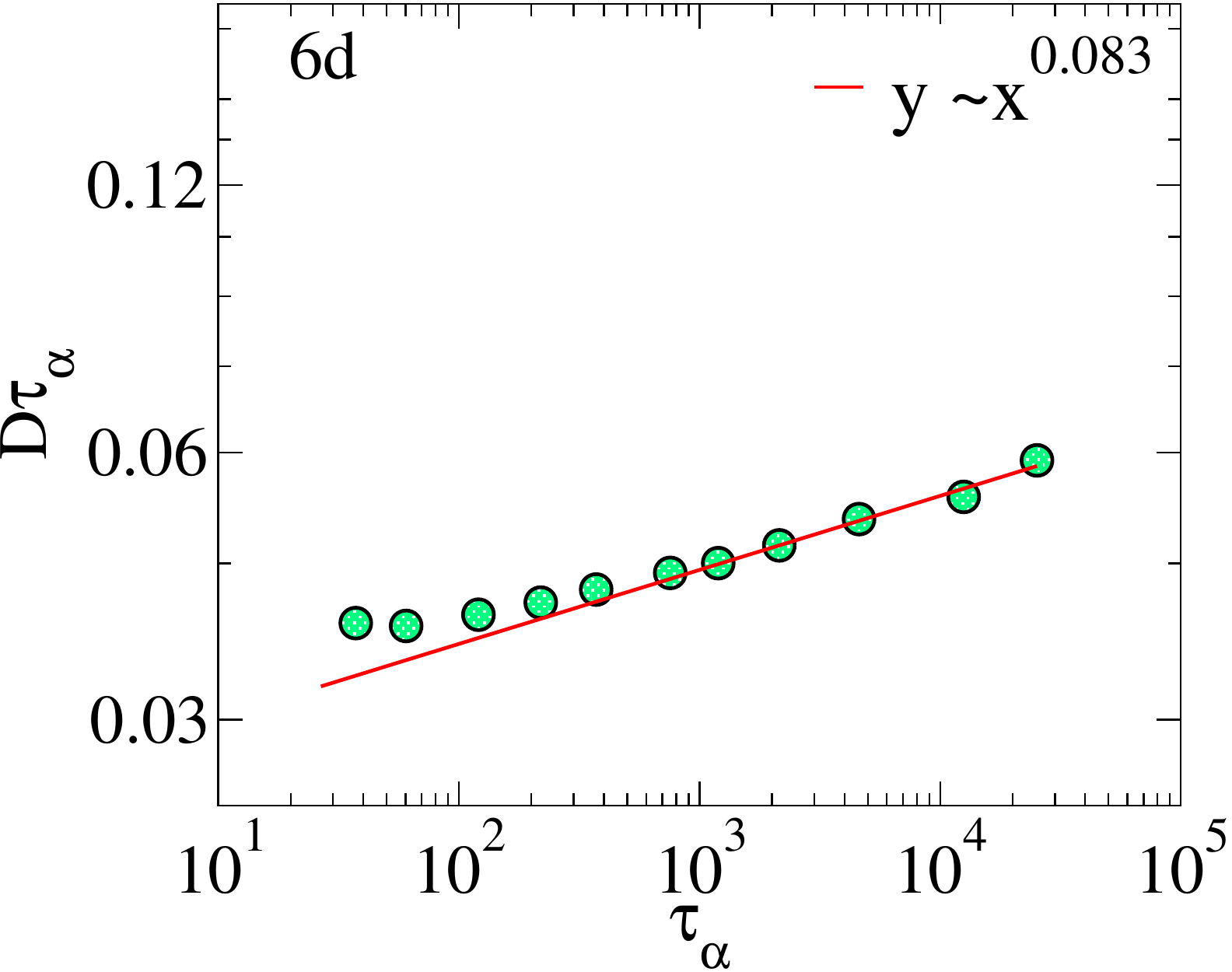}
\includegraphics[width=0.30\textwidth]{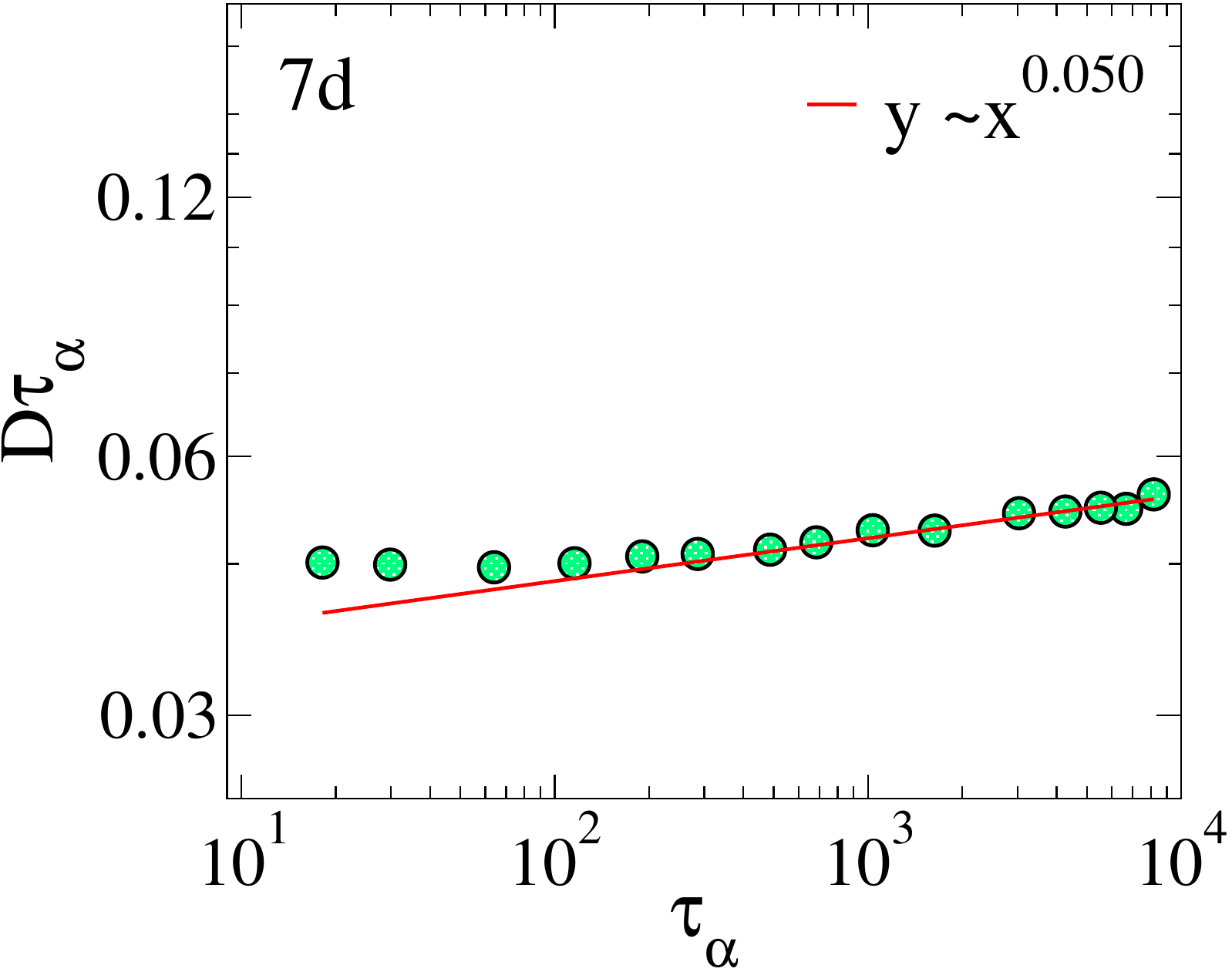}
\includegraphics[width=0.30\textwidth]{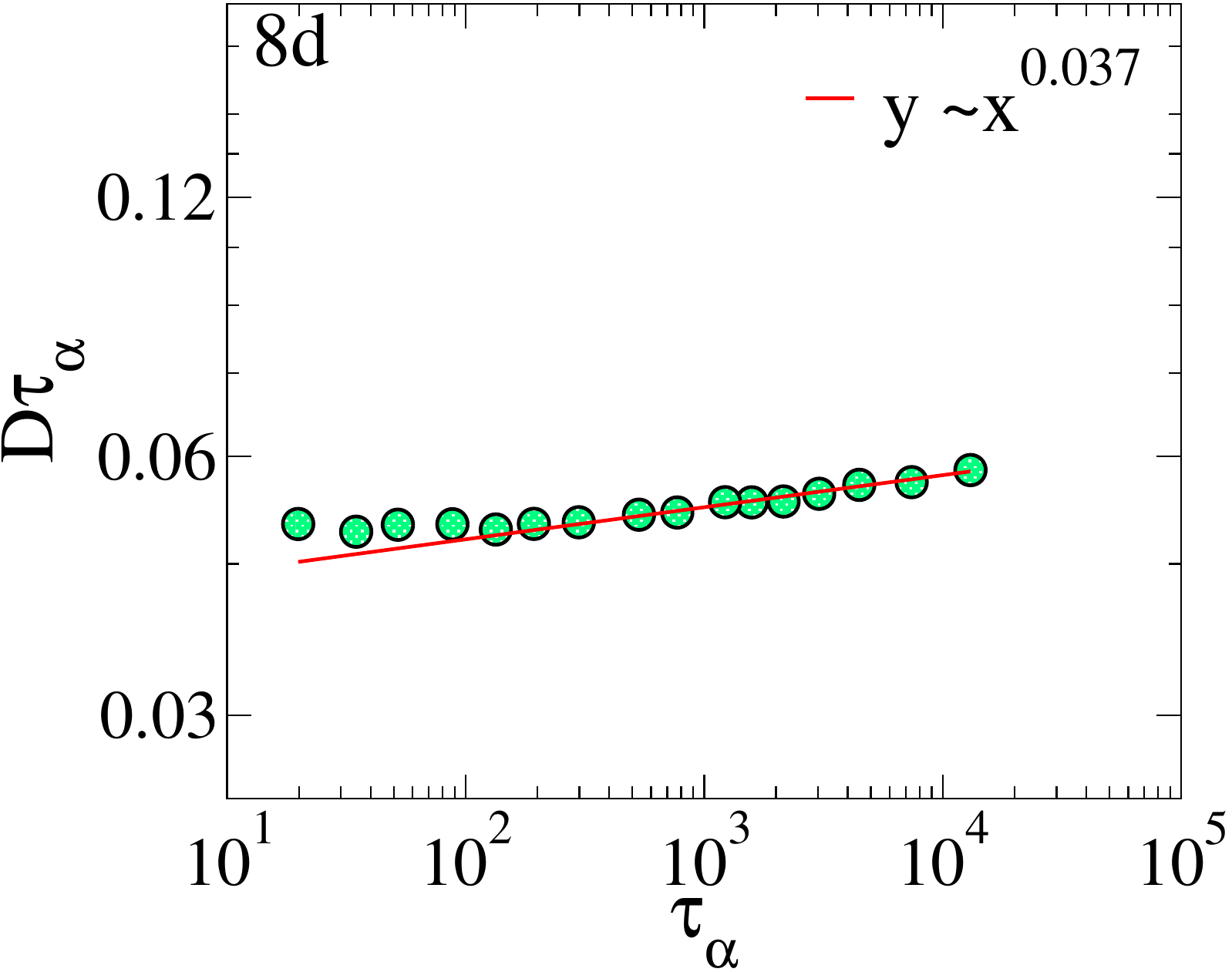}
\caption{$D \tau_{\alpha}$ is plotted against $\tau_{\alpha}$ in a log-log plot. The low temperature data is fitted with the form: $D \tau_{\alpha} \sim \tau_{\alpha}^{\omega}$. From the fit, we obtain $\omega$ for each spatial dimension.}
\label{fig-SER}
\end{figure}

\begin{figure}[htp]
\centering
\includegraphics[width=0.44\textwidth]{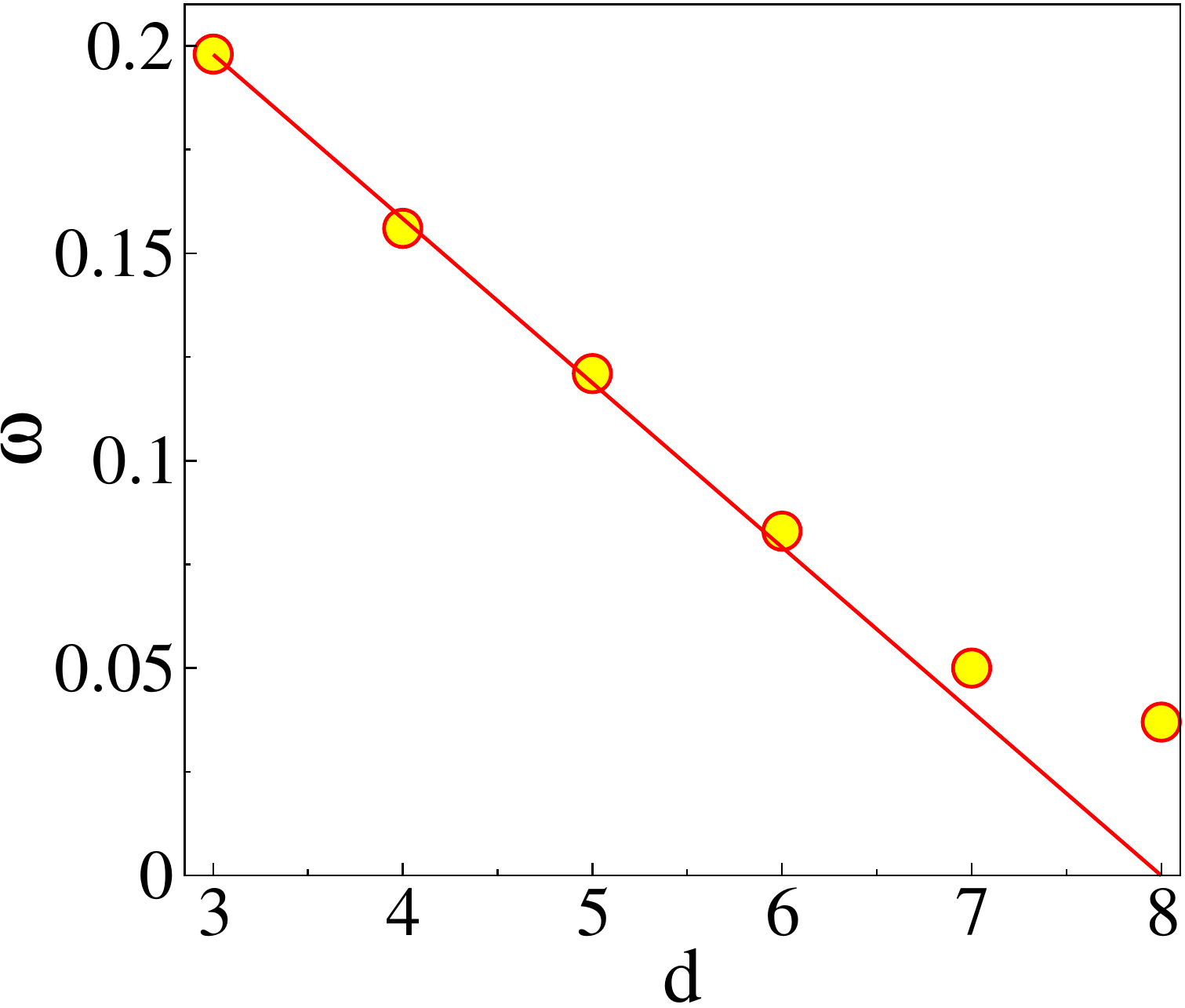}
\includegraphics[width=0.47\textwidth]{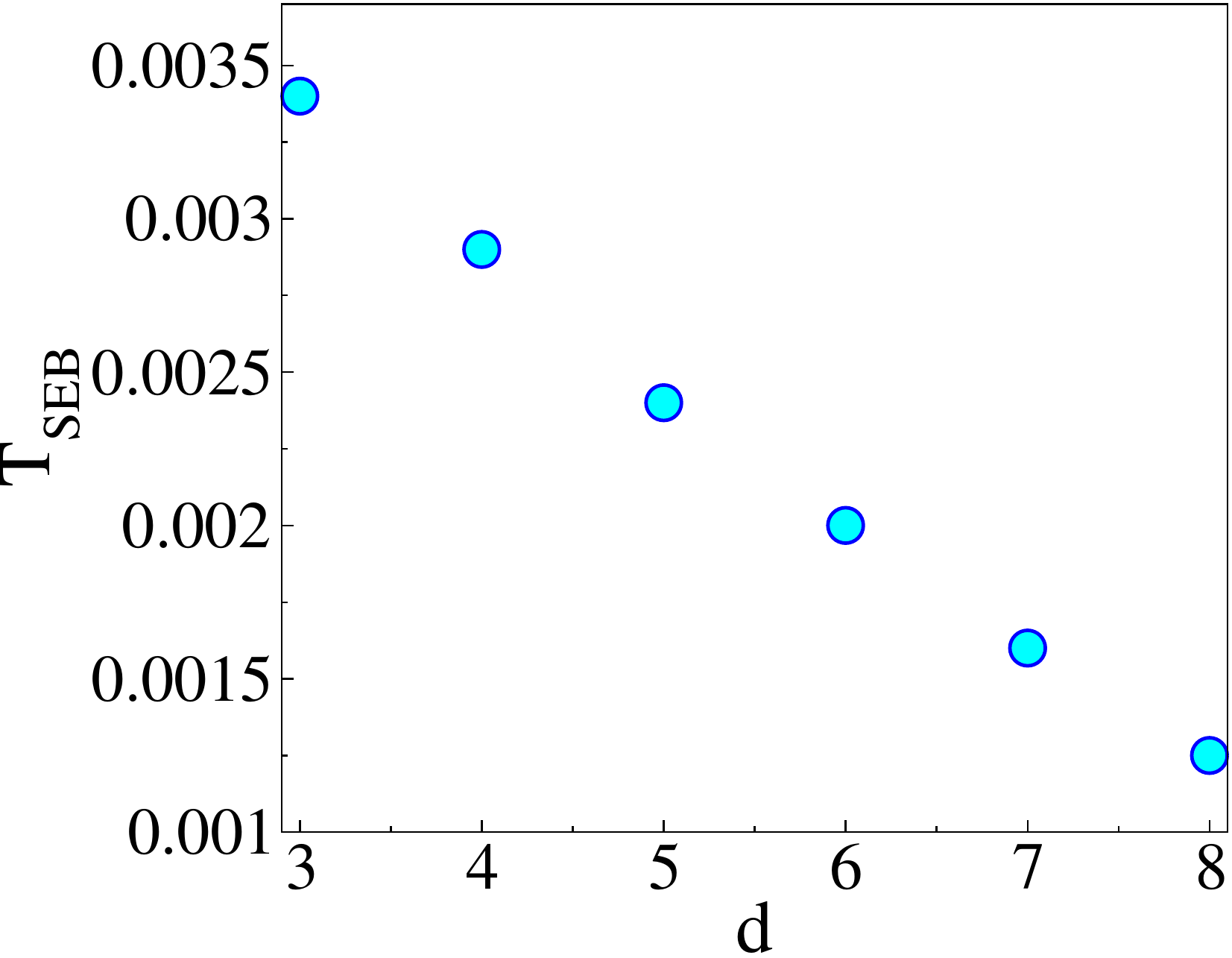}
\caption{Left panel: The exponent $\omega$ is plotted against the spatial dimensions. The exponent $\omega$ decreases with increasing spatial dimensionality. Right panel: $T_{SEB}$, the temperature where SER breaks down is plotted against spatial dimension. $T_{SEB}$ also decreases with increasing spatial dimensionality.}
\label{fig-omega}
\end{figure}

In Fig. \ref{fig-SER}, we show the diffusivity, $D$ multiplied by the relaxation time, $\tau_{\alpha}$, against $\tau_{\alpha}$ in log-log plot. We observe that $D \tau_{\alpha}$ is roughly constant for small $\tau_{\alpha}$ (high temperature), at least for $3d, 4d$ and $5d$, but display power law behaviour with a finite $\omega$ for large $\tau_{\alpha}$. We obtain the exponent $\omega$ by power-law fits of the form $D\tau_{\alpha} \sim \tau_{\alpha}^\omega$ for the data in the low temperature regime. In Fig. \ref{fig-omega} (Left panel), we show $\omega$ as a function of spatial dimension. We find that $\omega$ is large for $3$d, and decreases with increasing spatial dimensionality following a relation, $\omega \sim (d-d_c)$. We see $\omega$ becomes zero by extrapolation of the linear form at d$=8$, consistently with the idea that the upper critical dimension $d_u = 8$, and consistently with results for the hard sphere fluid \cite{charbonneau2012dimensional}. The numerical values of $\omega$ show a deviation from the linear fit for $d=7,8$ and do not vanish for $d = 8$, as also seen in \cite{charbonneau2012dimensional}. With the results available, we cannot further probe this issue. Improved numerical results and performing simulations at higher dimensions than $8$ will permit more precise statements in this regard. In Fig. \ref{fig-omega} (Right panel), we show the breakdown temperature for SER relation, $T_{SEB}$, as a function of spatial dimension. $T_{SEB}$ is defined as the temperature below which $D\tau_{\alpha}$ exceeds the high temperature value by $7.5 \%$..    

\subsection{Density Dependence and Comparison with Previous Results}

\begin{figure}[htp]
\centering
\includegraphics[width=0.38\textwidth]{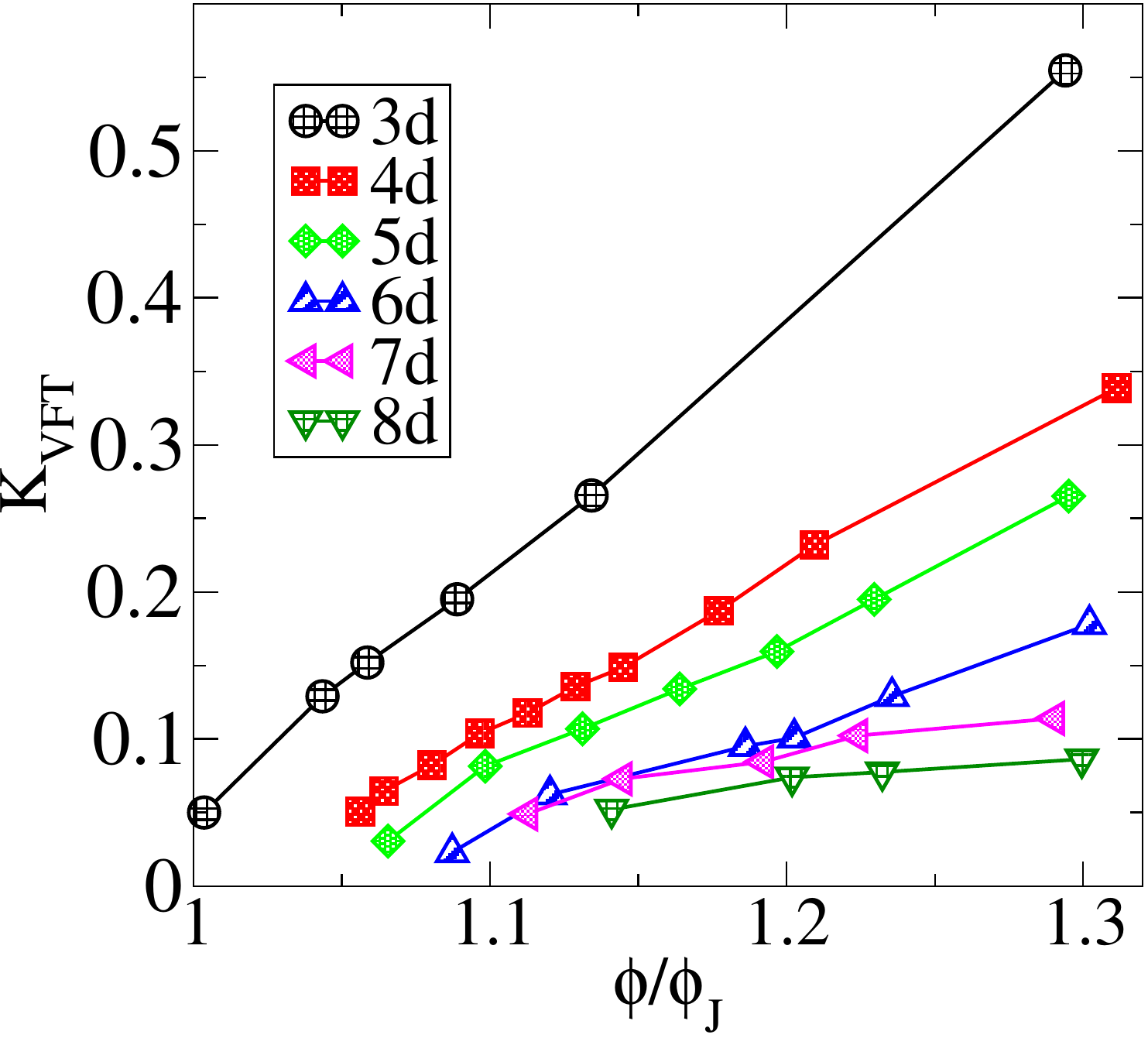}
\includegraphics[width=0.40\textwidth]{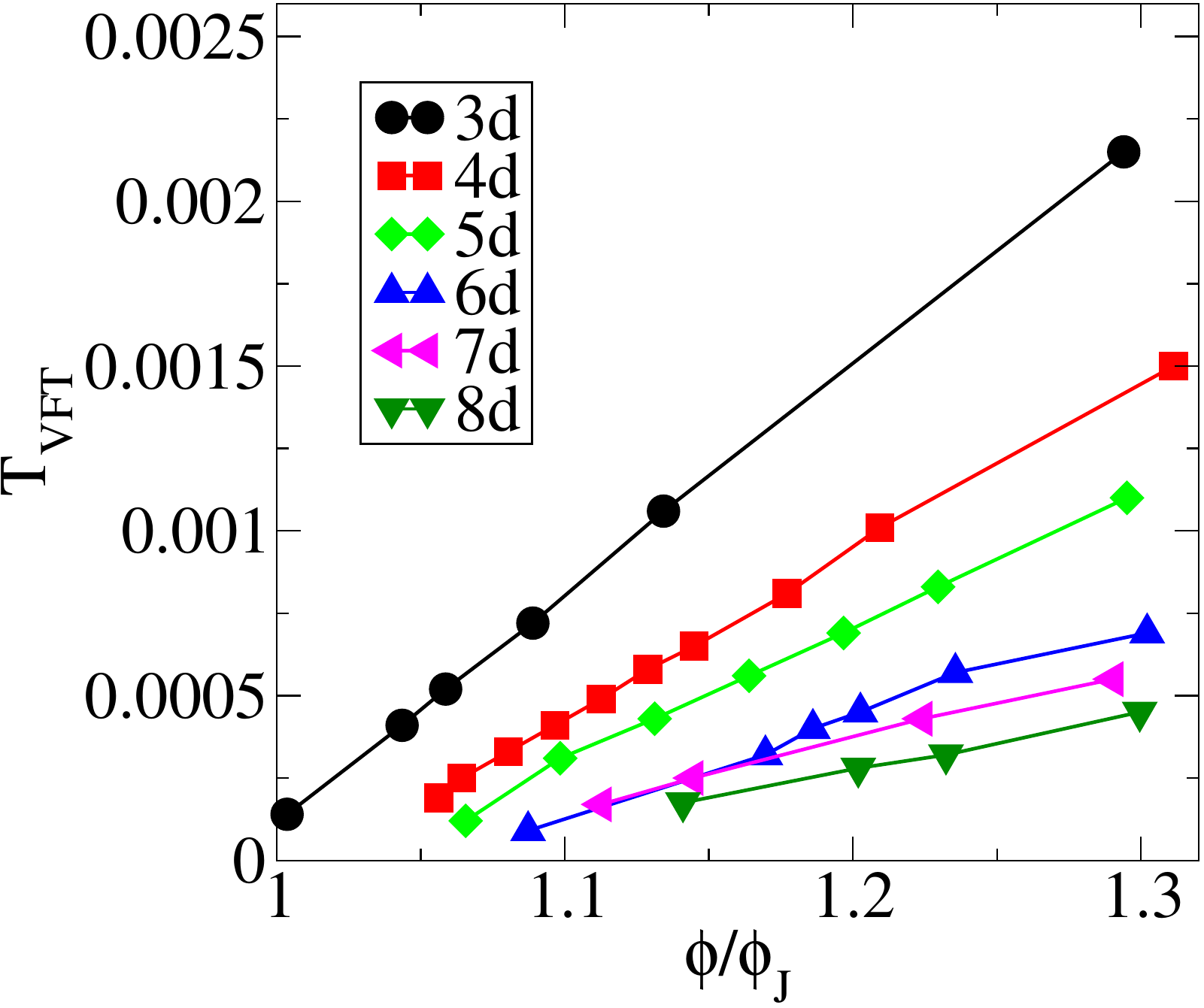}
\caption{Left panel: The kinetic fragility $K_{VFT}$ is shown as a function of scaled density $\phi/\phi_J$ for spatial dimensions $3-8$.  Right panel: $T_{VFT}$ is shown as a function of scaled density $\phi/\phi_J$ for spatial dimensions $3-8$.}
\label{scaled-VFT}
\end{figure}

We next briefly consider the dependence of the fragility on the density at which the liquids are studied, and address an apparent inconsistency with previous results. Further details of density dependence are discussed in detail in an accompanying paper. 
In Fig. \ref{scaled-VFT}, we show the kinetic fragility $K_{VFT}$ and the divergence temperature $T_{VFT}$ against density, scaled with $\phi_J$. We note that the kinetic agility $K_{VFT}$ decreases and nearly vanishes as the density is decreased towards $\phi_J$ (Indeed, for higher spatial dimensions, such vanishing appears to occur for densities higher than $\phi_J$, whose significance is discussed elsewhere), while at any fixed density, the fragility is a decreasing function of spatial dimensionality, consistently with the results discussed already for $\phi = 1.3 \phi_J$. Similarly to $K_{VFT}$, $T_{VFT}$ also decreases as the density is lowered, while being smaller for higher dimensions at fixed density. Thus, in comparing behaviour as a function of dimensionality, care must be exercised to compare results at the same scaled densities.

The observed dependence on density and spatial dimensionality helps explain an apparent inconsistency with results discussed by Sengupta \emph{et al.}\cite{sengupta2013breakdown}. 
In \cite{sengupta2013breakdown}, simulation results were shown for the Kob-Andersen (KA) binary Lennard-Jones mixture, and it was observed that liquids in $4d$ were less heterogeneous than in $3d$ 
(consistently with results here), but had larger fragility than in $3d$, which is not consistent with the present observations that the fragility too decreases with increasing spatial dimensionality.  

\begin{figure}[htp]
\centering
\includegraphics[scale=0.44]{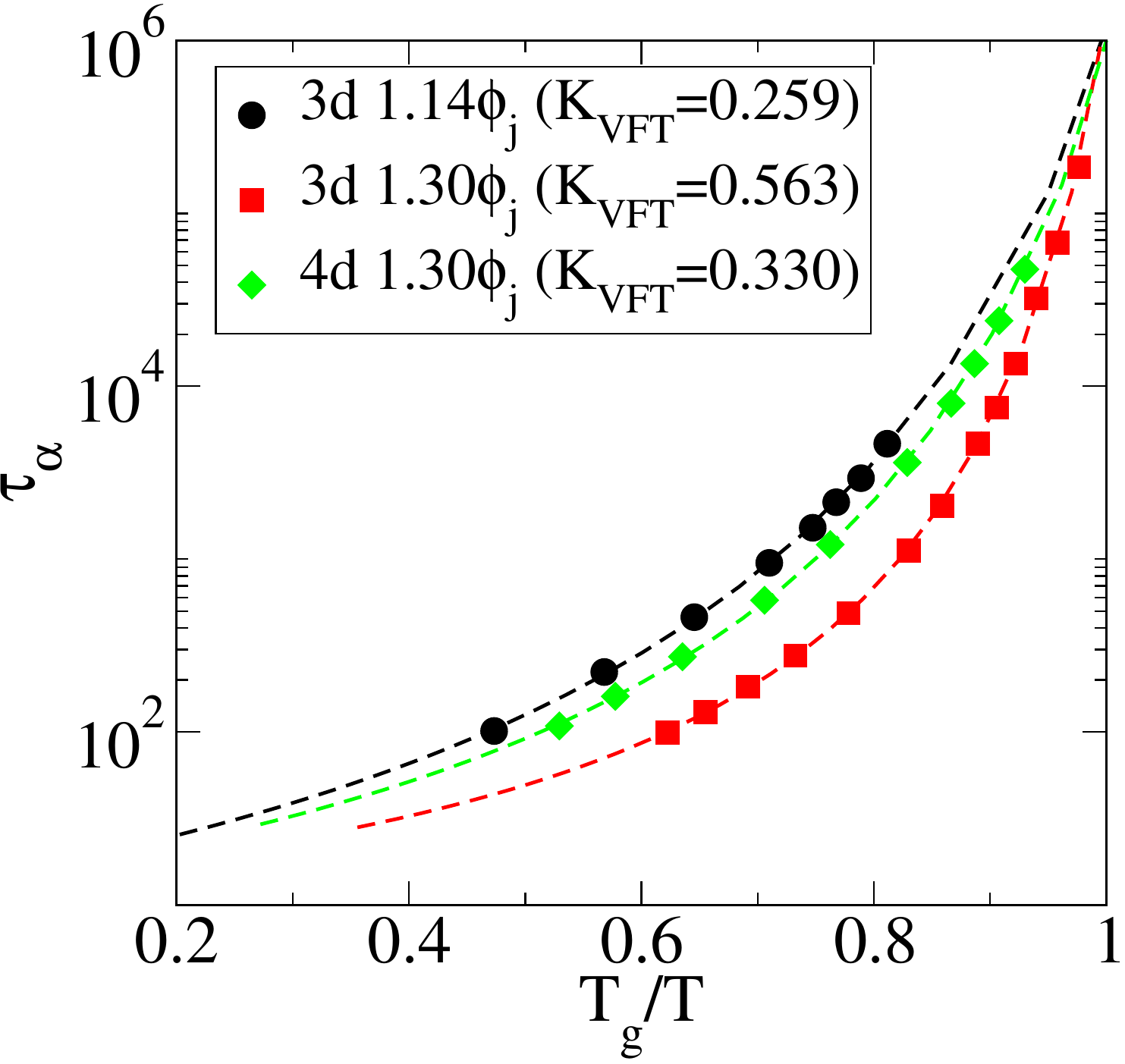}
\includegraphics[scale=0.45]{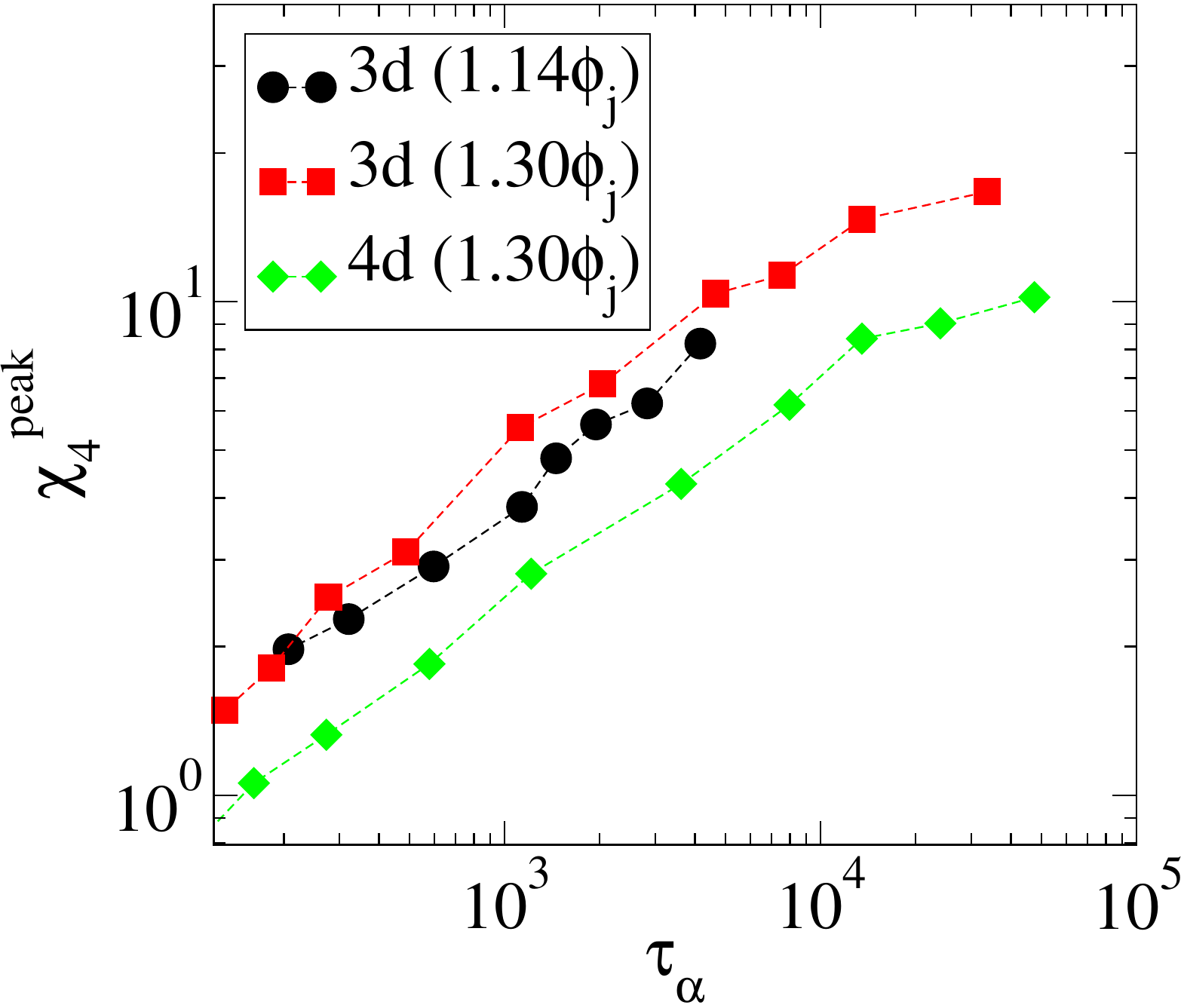}
\caption{Left panel: Relaxation times are shown in an Angell plot for (a) $3$d at two different densities, $1.14\phi_J$, $1.3\phi_J$ and (b) for $4$d at density $1.3\phi_J$. 
The kinetic fragility at $1.14\phi_J$ ($3$d) is lower than the $3d$ and $4d$ systems at density $1.3 \phi_J$. Right panel: $\chi_4^{peak}$ is plotted against $\tau_{\alpha}$ for $3$d at two different densities, $1.14\phi_J$ and $1.3\phi_J$ and for $4$d at $1.3\phi_J$. $\chi_4^{peak}$ values for $4$d at  $1.3\phi_J$ are lower than for the $3d$ systems at both  $1.14\phi_J$ and  $1.3\phi_J$.}
\label{shila-VFT}
\end{figure}

As noted above, the fragility as well as the heterogeneity depends upon the density for a given spatial dimension, and thus, to compare results in different dimensions, one must consider appropriate densities. For the $KA$ system, we do not have a jamming density to provide an appropriate scale, and instead, we use the zero temperature limit of the spinodal density for reference \cite{sastry2000PRL}. The spinodal density for the KA system is $\sim 1.1$ for $3$d whereas it is $\sim 1.4$ for $4$d. The simulations in \cite{sengupta2013breakdown} were performed at higher densities, $1.2$ for $3$d which is $1.09$ times the spinodal density whereas for $4$d the density of $1.6$ was employed, which is $1.14$ times the spinodal density. Thus, the scaled density in $4d$ is higher than the scaled density in $3d$, which leads to a higher fragility in the higher dimension. 
To illustrate this possibility, we consider for the soft sphere system a lower density of $1.14\phi_J$ in $3d$, in addition to $1.3\phi_J$, and compare with results in $4d$ at $1.3\phi_J$.

In Fig. \ref{shila-VFT} (Left panel), we show an Angell plot for three different cases: $1.14\phi_J$ and $1.30\phi_J$ at $3$d and $1.30\phi_J$ at $4$d. We note that the $3d$ system at  $1.14\phi_J$ has a lower fragility (as confirmed by calculating $K_{VFT}$) than the $4d$ system at  $1.3\phi_J$. 
Plotting $\chi_4^{peak}$ against $\tau_{\alpha}$ for the same three cases (Right panel, Fig. \ref{shila-VFT}), we see that the $4d$ system has the lowest heterogeneity. Thus, comparing the $3d$ system at   $1.14\phi_J$ with the $4d$ system at  $1.3\phi_J$ would lead to the conclusion that the heterogeneity in $3d$ is higher than in $4d$ while the fragility is higher in $4d$, whereas comparison at the same scaled density would lead to the conclusion that both the fragility and heterogeneity would decrease with increasing spatial dimensionality. We therefore conclude that the results in \cite{sengupta2013breakdown} can be understood consistently with our present results by noting the choice of densities in \cite{sengupta2013breakdown}. 

 


\section{Summary and Conclusions}
In conclusion, we have investigated the relationship between fragility, heterogeneity and the breakdown of the Stokes-Einstein relation in different spatial dimensions. Our results show that at fixed density, the fragility, the degree of heterogeneity and the degree of violation of the Stokes-Einstein relation decrease with increasing spatial dimensionality. The heterogeneity measures $\chi_4^{peak}$ and $\alpha_2^{peak}$ depend on the relaxation time $\tau_{\alpha}$ at high and moderate temperatures in a power law fashion, with power law exponents that do not depend on spatial dimensionality. The exponent $\omega$ that characterises the breakdown of the Stokes-Einstein relationship displays a nearly linear relationship with spatial dimensions that corresponds to a vanishing of $\omega$ at $d=8$, consistently with the idea that $d=8$ represents the upper critical dimension. The $\omega$ values in $d=7,8$ display small deviations from such linear behaviour with $d$, which requires further investigation including studies in dimensions above $8$. We show that fragilities decrease with density at all spatial dimensions, approaching Arrhenius behaviour close to the jamming density. The observed density dependence helps rationalise earlier results that suggested that fragility and heterogeneity may vary in opposite ways as a function of spatial dimensionality. 

\begin{acknowledgement}
We acknowledge the Thematic Unit of Excellence on Computational Materials Science, and the National Supercomputing Mission facility (Param Yukti) at the Jawaharlal Nehru Center for Advanced Scientific Research for computational resources. SK acknowledges support from Swarna Jayanti Fellowship grants DST/SJF/PSA-01/2018-19 and SB/SFJ/2019-20/05. SS acknowledges support through the JC Bose Fellowship  (JBR/2020/000015) SERB, DST (India).
\end{acknowledgement}

\bibliography{longDh}

\providecommand{\latin}[1]{#1}
\makeatletter
\providecommand{\doi}
  {\begingroup\let\do\@makeother\dospecials
  \catcode`\{=1 \catcode`\}=2 \doi@aux}
\providecommand{\doi@aux}[1]{\endgroup\texttt{#1}}
\makeatother
\providecommand*\mcitethebibliography{\thebibliography}
\csname @ifundefined\endcsname{endmcitethebibliography}
  {\let\endmcitethebibliography\endthebibliography}{}
\begin{mcitethebibliography}{64}
\providecommand*\natexlab[1]{#1}
\providecommand*\mciteSetBstSublistMode[1]{}
\providecommand*\mciteSetBstMaxWidthForm[2]{}
\providecommand*\mciteBstWouldAddEndPuncttrue
  {\def\EndOfBibitem{\unskip.}}
\providecommand*\mciteBstWouldAddEndPunctfalse
  {\let\EndOfBibitem\relax}
\providecommand*\mciteSetBstMidEndSepPunct[3]{}
\providecommand*\mciteSetBstSublistLabelBeginEnd[3]{}
\providecommand*\EndOfBibitem{}
\mciteSetBstSublistMode{f}
\mciteSetBstMaxWidthForm{subitem}{(\alph{mcitesubitemcount})}
\mciteSetBstSublistLabelBeginEnd
  {\mcitemaxwidthsubitemform\space}
  {\relax}
  {\relax}

\bibitem[Sillescu(1999)]{sillescu1999heterogeneity}
Sillescu,~H. \emph{Journal of Non-Crystalline Solids} \textbf{1999},
  \emph{243}, 81--108\relax
\mciteBstWouldAddEndPuncttrue
\mciteSetBstMidEndSepPunct{\mcitedefaultmidpunct}
{\mcitedefaultendpunct}{\mcitedefaultseppunct}\relax
\EndOfBibitem
\bibitem[Ediger(2000)]{ediger2000spatially}
Ediger,~M.~D. \emph{Annual review of physical chemistry} \textbf{2000},
  \emph{51}, 99--128\relax
\mciteBstWouldAddEndPuncttrue
\mciteSetBstMidEndSepPunct{\mcitedefaultmidpunct}
{\mcitedefaultendpunct}{\mcitedefaultseppunct}\relax
\EndOfBibitem
\bibitem[Berthier \latin{et~al.}(2011)Berthier, Biroli, Bouchaud, Cipelletti,
  and van Saarloos]{berthier2011dynamical}
Berthier,~L.; Biroli,~G.; Bouchaud,~J.-P.; Cipelletti,~L.; van Saarloos,~W.
  \emph{Dynamical heterogeneities in glasses, colloids, and granular media};
  OUP Oxford, 2011; Vol. 150\relax
\mciteBstWouldAddEndPuncttrue
\mciteSetBstMidEndSepPunct{\mcitedefaultmidpunct}
{\mcitedefaultendpunct}{\mcitedefaultseppunct}\relax
\EndOfBibitem
\bibitem[Kob \latin{et~al.}(1997)Kob, Donati, Plimpton, Poole, and
  Glotzer]{kob1997dynamical}
Kob,~W.; Donati,~C.; Plimpton,~S.~J.; Poole,~P.~H.; Glotzer,~S.~C.
  \emph{Physical review letters} \textbf{1997}, \emph{79}, 2827\relax
\mciteBstWouldAddEndPuncttrue
\mciteSetBstMidEndSepPunct{\mcitedefaultmidpunct}
{\mcitedefaultendpunct}{\mcitedefaultseppunct}\relax
\EndOfBibitem
\bibitem[Perera and Harrowell(1998)Perera, and Harrowell]{perera1998origin}
Perera,~D.~N.; Harrowell,~P. \emph{Physical review letters} \textbf{1998},
  \emph{81}, 120\relax
\mciteBstWouldAddEndPuncttrue
\mciteSetBstMidEndSepPunct{\mcitedefaultmidpunct}
{\mcitedefaultendpunct}{\mcitedefaultseppunct}\relax
\EndOfBibitem
\bibitem[Yamamoto and Onuki(1998)Yamamoto, and
  Onuki]{yamamoto1998heterogeneous}
Yamamoto,~R.; Onuki,~A. \emph{Physical review letters} \textbf{1998},
  \emph{81}, 4915\relax
\mciteBstWouldAddEndPuncttrue
\mciteSetBstMidEndSepPunct{\mcitedefaultmidpunct}
{\mcitedefaultendpunct}{\mcitedefaultseppunct}\relax
\EndOfBibitem
\bibitem[Donati \latin{et~al.}(2002)Donati, Franz, Glotzer, and
  Parisi]{donati2002theory}
Donati,~C.; Franz,~S.; Glotzer,~S.~C.; Parisi,~G. \emph{Journal of
  non-crystalline solids} \textbf{2002}, \emph{307}, 215--224\relax
\mciteBstWouldAddEndPuncttrue
\mciteSetBstMidEndSepPunct{\mcitedefaultmidpunct}
{\mcitedefaultendpunct}{\mcitedefaultseppunct}\relax
\EndOfBibitem
\bibitem[Flenner \latin{et~al.}(2014)Flenner, Staley, and
  Szamel]{flenner2014universal}
Flenner,~E.; Staley,~H.; Szamel,~G. \emph{Physical review letters}
  \textbf{2014}, \emph{112}, 097801\relax
\mciteBstWouldAddEndPuncttrue
\mciteSetBstMidEndSepPunct{\mcitedefaultmidpunct}
{\mcitedefaultendpunct}{\mcitedefaultseppunct}\relax
\EndOfBibitem
\bibitem[Karmakar \latin{et~al.}(2014)Karmakar, Dasgupta, and
  Sastry]{karmakar2014growing}
Karmakar,~S.; Dasgupta,~C.; Sastry,~S. \emph{Annu. Rev. Condens. Matter Phys.}
  \textbf{2014}, \emph{5}, 255--284\relax
\mciteBstWouldAddEndPuncttrue
\mciteSetBstMidEndSepPunct{\mcitedefaultmidpunct}
{\mcitedefaultendpunct}{\mcitedefaultseppunct}\relax
\EndOfBibitem
\bibitem[Karmakar \latin{et~al.}(2016)Karmakar, Dasgupta, and
  Sastry]{karmakar2016length}
Karmakar,~S.; Dasgupta,~C.; Sastry,~S. \emph{Reports on Progress in Physics}
  \textbf{2016}, \emph{79}, 016601\relax
\mciteBstWouldAddEndPuncttrue
\mciteSetBstMidEndSepPunct{\mcitedefaultmidpunct}
{\mcitedefaultendpunct}{\mcitedefaultseppunct}\relax
\EndOfBibitem
\bibitem[R\"ossler(1990)]{Rossler1990}
R\"ossler,~E. \emph{Phys. Rev. Lett.} \textbf{1990}, \emph{65},
  1595--1598\relax
\mciteBstWouldAddEndPuncttrue
\mciteSetBstMidEndSepPunct{\mcitedefaultmidpunct}
{\mcitedefaultendpunct}{\mcitedefaultseppunct}\relax
\EndOfBibitem
\bibitem[Fujara \latin{et~al.}(1992)Fujara, Geil, Sillescu, and
  Fleischer]{fujara1992translational}
Fujara,~F.; Geil,~B.; Sillescu,~H.; Fleischer,~G. \emph{Zeitschrift f{\"u}r
  Physik B Condensed Matter} \textbf{1992}, \emph{88}, 195--204\relax
\mciteBstWouldAddEndPuncttrue
\mciteSetBstMidEndSepPunct{\mcitedefaultmidpunct}
{\mcitedefaultendpunct}{\mcitedefaultseppunct}\relax
\EndOfBibitem
\bibitem[Thirumalai and Mountain(1993)Thirumalai, and
  Mountain]{thirumalai1993activated}
Thirumalai,~D.; Mountain,~R.~D. \emph{Physical Review E} \textbf{1993},
  \emph{47}, 479\relax
\mciteBstWouldAddEndPuncttrue
\mciteSetBstMidEndSepPunct{\mcitedefaultmidpunct}
{\mcitedefaultendpunct}{\mcitedefaultseppunct}\relax
\EndOfBibitem
\bibitem[Stillinger and Hodgdon(1994)Stillinger, and
  Hodgdon]{stillinger1994translation}
Stillinger,~F.~H.; Hodgdon,~J.~A. \emph{Physical review E} \textbf{1994},
  \emph{50}, 2064\relax
\mciteBstWouldAddEndPuncttrue
\mciteSetBstMidEndSepPunct{\mcitedefaultmidpunct}
{\mcitedefaultendpunct}{\mcitedefaultseppunct}\relax
\EndOfBibitem
\bibitem[Tarjus and Kivelson(1995)Tarjus, and Kivelson]{tarjus1995breakdown}
Tarjus,~G.; Kivelson,~D. \emph{The Journal of chemical physics} \textbf{1995},
  \emph{103}, 3071--3073\relax
\mciteBstWouldAddEndPuncttrue
\mciteSetBstMidEndSepPunct{\mcitedefaultmidpunct}
{\mcitedefaultendpunct}{\mcitedefaultseppunct}\relax
\EndOfBibitem
\bibitem[Cicerone and Ediger(1996)Cicerone, and Ediger]{cicerone1996enhanced}
Cicerone,~M.~T.; Ediger,~M.~D. \emph{The Journal of chemical physics}
  \textbf{1996}, \emph{104}, 7210--7218\relax
\mciteBstWouldAddEndPuncttrue
\mciteSetBstMidEndSepPunct{\mcitedefaultmidpunct}
{\mcitedefaultendpunct}{\mcitedefaultseppunct}\relax
\EndOfBibitem
\bibitem[Berthier \latin{et~al.}(2004)Berthier, Chandler, and
  Garrahan]{berthier2004length}
Berthier,~L.; Chandler,~D.; Garrahan,~J.~P. \emph{EPL (Europhysics Letters)}
  \textbf{2004}, \emph{69}, 320\relax
\mciteBstWouldAddEndPuncttrue
\mciteSetBstMidEndSepPunct{\mcitedefaultmidpunct}
{\mcitedefaultendpunct}{\mcitedefaultseppunct}\relax
\EndOfBibitem
\bibitem[Berthier(2004)]{berthier2004time}
Berthier,~L. \emph{Physical Review E} \textbf{2004}, \emph{69}, 020201\relax
\mciteBstWouldAddEndPuncttrue
\mciteSetBstMidEndSepPunct{\mcitedefaultmidpunct}
{\mcitedefaultendpunct}{\mcitedefaultseppunct}\relax
\EndOfBibitem
\bibitem[Jung \latin{et~al.}(2004)Jung, Garrahan, and Chandler]{Jung2004}
Jung,~Y.; Garrahan,~J.~P.; Chandler,~D. \emph{Phys. Rev. E} \textbf{2004},
  \emph{69}, 061205\relax
\mciteBstWouldAddEndPuncttrue
\mciteSetBstMidEndSepPunct{\mcitedefaultmidpunct}
{\mcitedefaultendpunct}{\mcitedefaultseppunct}\relax
\EndOfBibitem
\bibitem[Kim and Keyes(2005)Kim, and Keyes]{kim2005breakdown}
Kim,~J.; Keyes,~T. \emph{The Journal of Physical Chemistry B} \textbf{2005},
  \emph{109}, 21445--21448\relax
\mciteBstWouldAddEndPuncttrue
\mciteSetBstMidEndSepPunct{\mcitedefaultmidpunct}
{\mcitedefaultendpunct}{\mcitedefaultseppunct}\relax
\EndOfBibitem
\bibitem[Kumar \latin{et~al.}(2006)Kumar, Szamel, and Douglas]{kumar2006}
Kumar,~S.~K.; Szamel,~G.; Douglas,~J.~F. \emph{The Journal of chemical physics}
  \textbf{2006}, \emph{124}, 214501\relax
\mciteBstWouldAddEndPuncttrue
\mciteSetBstMidEndSepPunct{\mcitedefaultmidpunct}
{\mcitedefaultendpunct}{\mcitedefaultseppunct}\relax
\EndOfBibitem
\bibitem[Becker \latin{et~al.}(2006)Becker, Poole, and
  Starr]{becker2006fractional}
Becker,~S.~R.; Poole,~P.~H.; Starr,~F.~W. \emph{Physical review letters}
  \textbf{2006}, \emph{97}, 055901\relax
\mciteBstWouldAddEndPuncttrue
\mciteSetBstMidEndSepPunct{\mcitedefaultmidpunct}
{\mcitedefaultendpunct}{\mcitedefaultseppunct}\relax
\EndOfBibitem
\bibitem[Chong(2008)]{chong2008connections}
Chong,~S.-H. \emph{Physical Review E} \textbf{2008}, \emph{78}, 041501\relax
\mciteBstWouldAddEndPuncttrue
\mciteSetBstMidEndSepPunct{\mcitedefaultmidpunct}
{\mcitedefaultendpunct}{\mcitedefaultseppunct}\relax
\EndOfBibitem
\bibitem[Chong and Kob(2009)Chong, and Kob]{chong2009coupling}
Chong,~S.-H.; Kob,~W. \emph{Physical review letters} \textbf{2009}, \emph{102},
  025702\relax
\mciteBstWouldAddEndPuncttrue
\mciteSetBstMidEndSepPunct{\mcitedefaultmidpunct}
{\mcitedefaultendpunct}{\mcitedefaultseppunct}\relax
\EndOfBibitem
\bibitem[Sengupta \latin{et~al.}(2013)Sengupta, Karmakar, Dasgupta, and
  Sastry]{sengupta2013breakdown}
Sengupta,~S.; Karmakar,~S.; Dasgupta,~C.; Sastry,~S. \emph{The Journal of
  chemical physics} \textbf{2013}, \emph{138}, 12A548\relax
\mciteBstWouldAddEndPuncttrue
\mciteSetBstMidEndSepPunct{\mcitedefaultmidpunct}
{\mcitedefaultendpunct}{\mcitedefaultseppunct}\relax
\EndOfBibitem
\bibitem[Sengupta and Karmakar(2014)Sengupta, and
  Karmakar]{sengupta2014distribution}
Sengupta,~S.; Karmakar,~S. \emph{The Journal of chemical physics}
  \textbf{2014}, \emph{140}, 224505\relax
\mciteBstWouldAddEndPuncttrue
\mciteSetBstMidEndSepPunct{\mcitedefaultmidpunct}
{\mcitedefaultendpunct}{\mcitedefaultseppunct}\relax
\EndOfBibitem
\bibitem[Charbonneau \latin{et~al.}(2013)Charbonneau, Charbonneau, Jin, Parisi,
  and Zamponi]{charbonneau2013dimensional}
Charbonneau,~B.; Charbonneau,~P.; Jin,~Y.; Parisi,~G.; Zamponi,~F. \emph{The
  Journal of chemical physics} \textbf{2013}, \emph{139}, 164502\relax
\mciteBstWouldAddEndPuncttrue
\mciteSetBstMidEndSepPunct{\mcitedefaultmidpunct}
{\mcitedefaultendpunct}{\mcitedefaultseppunct}\relax
\EndOfBibitem
\bibitem[Charbonneau \latin{et~al.}(2014)Charbonneau, Jin, Parisi, and
  Zamponi]{charbonneau2014hopping}
Charbonneau,~P.; Jin,~Y.; Parisi,~G.; Zamponi,~F. \emph{Proceedings of the
  National Academy of Sciences} \textbf{2014}, \emph{111}, 15025--15030\relax
\mciteBstWouldAddEndPuncttrue
\mciteSetBstMidEndSepPunct{\mcitedefaultmidpunct}
{\mcitedefaultendpunct}{\mcitedefaultseppunct}\relax
\EndOfBibitem
\bibitem[Parmar \latin{et~al.}(2017)Parmar, Sengupta, and
  Sastry]{parmar2017length}
Parmar,~A.~D.; Sengupta,~S.; Sastry,~S. \emph{Physical review letters}
  \textbf{2017}, \emph{119}, 056001\relax
\mciteBstWouldAddEndPuncttrue
\mciteSetBstMidEndSepPunct{\mcitedefaultmidpunct}
{\mcitedefaultendpunct}{\mcitedefaultseppunct}\relax
\EndOfBibitem
\bibitem[Nandi and Bhattacharyya(2019)Nandi, and
  Bhattacharyya]{nandi2019continuous}
Nandi,~M.~K.; Bhattacharyya,~S.~M. \emph{Journal of Physics: Condensed Matter}
  \textbf{2019}, \emph{32}, 064001\relax
\mciteBstWouldAddEndPuncttrue
\mciteSetBstMidEndSepPunct{\mcitedefaultmidpunct}
{\mcitedefaultendpunct}{\mcitedefaultseppunct}\relax
\EndOfBibitem
\bibitem[B{\"o}hmer \latin{et~al.}(1993)B{\"o}hmer, Ngai, Angell, and
  Plazek]{bohmer1993nonexponential}
B{\"o}hmer,~R.; Ngai,~K.; Angell,~C.~A.; Plazek,~D. \emph{The Journal of
  chemical physics} \textbf{1993}, \emph{99}, 4201--4209\relax
\mciteBstWouldAddEndPuncttrue
\mciteSetBstMidEndSepPunct{\mcitedefaultmidpunct}
{\mcitedefaultendpunct}{\mcitedefaultseppunct}\relax
\EndOfBibitem
\bibitem[Angell(1991)]{angell1991relaxation}
Angell,~C. \emph{Journal of Non-Crystalline Solids} \textbf{1991}, \emph{131},
  13--31\relax
\mciteBstWouldAddEndPuncttrue
\mciteSetBstMidEndSepPunct{\mcitedefaultmidpunct}
{\mcitedefaultendpunct}{\mcitedefaultseppunct}\relax
\EndOfBibitem
\bibitem[Nielsen \latin{et~al.}(2009)Nielsen, Christensen, Jakobsen, Niss,
  Olsen, Richert, and Dyre]{Dyre2009}
Nielsen,~A.~I.; Christensen,~T.; Jakobsen,~B.; Niss,~K.; Olsen,~N.~B.;
  Richert,~R.; Dyre,~J.~C. \emph{The Journal of chemical physics}
  \textbf{2009}, \emph{130}, 154508\relax
\mciteBstWouldAddEndPuncttrue
\mciteSetBstMidEndSepPunct{\mcitedefaultmidpunct}
{\mcitedefaultendpunct}{\mcitedefaultseppunct}\relax
\EndOfBibitem
\bibitem[Douglas \latin{et~al.}(2006)Douglas, Dudowicz, and
  Freed]{douglas2006does}
Douglas,~J.~F.; Dudowicz,~J.; Freed,~K.~F. \emph{The Journal of chemical
  physics} \textbf{2006}, \emph{125}, 144907\relax
\mciteBstWouldAddEndPuncttrue
\mciteSetBstMidEndSepPunct{\mcitedefaultmidpunct}
{\mcitedefaultendpunct}{\mcitedefaultseppunct}\relax
\EndOfBibitem
\bibitem[Dudowicz \latin{et~al.}(2007)Dudowicz, Freed, and
  Douglas]{dudowicz2007}
Dudowicz,~J.; Freed,~K.~F.; Douglas,~J.~F. \emph{Advances in Chemical Physics};
  John Wiley \& Sons, Ltd, 2007; Chapter 3, pp 125--222\relax
\mciteBstWouldAddEndPuncttrue
\mciteSetBstMidEndSepPunct{\mcitedefaultmidpunct}
{\mcitedefaultendpunct}{\mcitedefaultseppunct}\relax
\EndOfBibitem
\bibitem[Starr \latin{et~al.}(2013)Starr, Douglas, and
  Sastry]{starr2013relationship}
Starr,~F.~W.; Douglas,~J.~F.; Sastry,~S. \emph{The Journal of chemical physics}
  \textbf{2013}, \emph{138}, 12A541\relax
\mciteBstWouldAddEndPuncttrue
\mciteSetBstMidEndSepPunct{\mcitedefaultmidpunct}
{\mcitedefaultendpunct}{\mcitedefaultseppunct}\relax
\EndOfBibitem
\bibitem[Betancourt \latin{et~al.}(2013)Betancourt, Douglas, and
  Starr]{betancourt2013fragility}
Betancourt,~B. A.~P.; Douglas,~J.~F.; Starr,~F.~W. \emph{Soft Matter}
  \textbf{2013}, \emph{9}, 241--254\relax
\mciteBstWouldAddEndPuncttrue
\mciteSetBstMidEndSepPunct{\mcitedefaultmidpunct}
{\mcitedefaultendpunct}{\mcitedefaultseppunct}\relax
\EndOfBibitem
\bibitem[Kirkpatrick and Thirumalai(1987)Kirkpatrick, and
  Thirumalai]{kirkpatrick1987dynamics}
Kirkpatrick,~T.~R.; Thirumalai,~D. \emph{Physical review letters}
  \textbf{1987}, \emph{58}, 2091\relax
\mciteBstWouldAddEndPuncttrue
\mciteSetBstMidEndSepPunct{\mcitedefaultmidpunct}
{\mcitedefaultendpunct}{\mcitedefaultseppunct}\relax
\EndOfBibitem
\bibitem[Kirkpatrick and Thirumalai(1988)Kirkpatrick, and
  Thirumalai]{kirkpatrick1988comparison}
Kirkpatrick,~T.~R.; Thirumalai,~D. \emph{Physical Review A} \textbf{1988},
  \emph{37}, 4439\relax
\mciteBstWouldAddEndPuncttrue
\mciteSetBstMidEndSepPunct{\mcitedefaultmidpunct}
{\mcitedefaultendpunct}{\mcitedefaultseppunct}\relax
\EndOfBibitem
\bibitem[Kirkpatrick \latin{et~al.}(1989)Kirkpatrick, Thirumalai, and
  Wolynes]{kirkpatrick1989scaling}
Kirkpatrick,~T.~R.; Thirumalai,~D.; Wolynes,~P.~G. \emph{Physical Review A}
  \textbf{1989}, \emph{40}, 1045\relax
\mciteBstWouldAddEndPuncttrue
\mciteSetBstMidEndSepPunct{\mcitedefaultmidpunct}
{\mcitedefaultendpunct}{\mcitedefaultseppunct}\relax
\EndOfBibitem
\bibitem[Biroli \latin{et~al.}(2006)Biroli, Bouchaud, Miyazaki, and
  Reichman]{biroli2006inhomogeneous}
Biroli,~G.; Bouchaud,~J.-P.; Miyazaki,~K.; Reichman,~D.~R. \emph{Physical
  review letters} \textbf{2006}, \emph{97}, 195701\relax
\mciteBstWouldAddEndPuncttrue
\mciteSetBstMidEndSepPunct{\mcitedefaultmidpunct}
{\mcitedefaultendpunct}{\mcitedefaultseppunct}\relax
\EndOfBibitem
\bibitem[Charbonneau \latin{et~al.}(2017)Charbonneau, Kurchan, Parisi, Urbani,
  and Zamponi]{charbonneau2017glass}
Charbonneau,~P.; Kurchan,~J.; Parisi,~G.; Urbani,~P.; Zamponi,~F. \emph{Annual
  Review of Condensed Matter Physics} \textbf{2017}, \emph{8}, 265--288\relax
\mciteBstWouldAddEndPuncttrue
\mciteSetBstMidEndSepPunct{\mcitedefaultmidpunct}
{\mcitedefaultendpunct}{\mcitedefaultseppunct}\relax
\EndOfBibitem
\bibitem[Manacorda \latin{et~al.}(2020)Manacorda, Schehr, and
  Zamponi]{Manacorda2020}
Manacorda,~A.; Schehr,~G.; Zamponi,~F. \emph{The Journal of Chemical Physics}
  \textbf{2020}, \emph{152}, 164506\relax
\mciteBstWouldAddEndPuncttrue
\mciteSetBstMidEndSepPunct{\mcitedefaultmidpunct}
{\mcitedefaultendpunct}{\mcitedefaultseppunct}\relax
\EndOfBibitem
\bibitem[Biroli and Bouchaud(2007)Biroli, and Bouchaud]{Biroli_2007}
Biroli,~G.; Bouchaud,~J.-P. \emph{Journal of Physics: Condensed Matter}
  \textbf{2007}, \emph{19}, 205101\relax
\mciteBstWouldAddEndPuncttrue
\mciteSetBstMidEndSepPunct{\mcitedefaultmidpunct}
{\mcitedefaultendpunct}{\mcitedefaultseppunct}\relax
\EndOfBibitem
\bibitem[Franz \latin{et~al.}(2012)Franz, Jacquin, Parisi, Urbani, and
  Zamponi]{Franz18725}
Franz,~S.; Jacquin,~H.; Parisi,~G.; Urbani,~P.; Zamponi,~F. \emph{Proceedings
  of the National Academy of Sciences} \textbf{2012}, \emph{109},
  18725--18730\relax
\mciteBstWouldAddEndPuncttrue
\mciteSetBstMidEndSepPunct{\mcitedefaultmidpunct}
{\mcitedefaultendpunct}{\mcitedefaultseppunct}\relax
\EndOfBibitem
\bibitem[Xu \latin{et~al.}(2016)Xu, Douglas, and Freed]{xu2016entropy}
Xu,~W.-S.; Douglas,~J.~F.; Freed,~K.~F. \emph{Adv. Chem. Phys} \textbf{2016},
  \emph{161}, 443--497\relax
\mciteBstWouldAddEndPuncttrue
\mciteSetBstMidEndSepPunct{\mcitedefaultmidpunct}
{\mcitedefaultendpunct}{\mcitedefaultseppunct}\relax
\EndOfBibitem
\bibitem[Eaves and Reichman(2009)Eaves, and Reichman]{eaves2009spatial}
Eaves,~J.~D.; Reichman,~D.~R. \emph{Proceedings of the National Academy of
  Sciences} \textbf{2009}, \emph{106}, 15171--15175\relax
\mciteBstWouldAddEndPuncttrue
\mciteSetBstMidEndSepPunct{\mcitedefaultmidpunct}
{\mcitedefaultendpunct}{\mcitedefaultseppunct}\relax
\EndOfBibitem
\bibitem[Durian(1995)]{durian1995foam}
Durian,~D.~J. \emph{Physical review letters} \textbf{1995}, \emph{75},
  4780\relax
\mciteBstWouldAddEndPuncttrue
\mciteSetBstMidEndSepPunct{\mcitedefaultmidpunct}
{\mcitedefaultendpunct}{\mcitedefaultseppunct}\relax
\EndOfBibitem
\bibitem[Berthier and Witten(2009)Berthier, and
  Witten]{berthier2009compressing}
Berthier,~L.; Witten,~T.~A. \emph{EPL (Europhysics Letters)} \textbf{2009},
  \emph{86}, 10001\relax
\mciteBstWouldAddEndPuncttrue
\mciteSetBstMidEndSepPunct{\mcitedefaultmidpunct}
{\mcitedefaultendpunct}{\mcitedefaultseppunct}\relax
\EndOfBibitem
\bibitem[Charbonneau \latin{et~al.}(2011)Charbonneau, Ikeda, Parisi, and
  Zamponi]{charbonneau2011glass}
Charbonneau,~P.; Ikeda,~A.; Parisi,~G.; Zamponi,~F. \emph{Physical review
  letters} \textbf{2011}, \emph{107}, 185702\relax
\mciteBstWouldAddEndPuncttrue
\mciteSetBstMidEndSepPunct{\mcitedefaultmidpunct}
{\mcitedefaultendpunct}{\mcitedefaultseppunct}\relax
\EndOfBibitem
\bibitem[Brown and Clarke(1984)Brown, and Clarke]{brown1984comparison}
Brown,~D.; Clarke,~J. \emph{Molecular Physics} \textbf{1984}, \emph{51},
  1243--1252\relax
\mciteBstWouldAddEndPuncttrue
\mciteSetBstMidEndSepPunct{\mcitedefaultmidpunct}
{\mcitedefaultendpunct}{\mcitedefaultseppunct}\relax
\EndOfBibitem
\bibitem[La{\v{c}}evi{\'c} \latin{et~al.}(2003)La{\v{c}}evi{\'c}, Starr,
  Schr{\o}der, and Glotzer]{lavcevic2003spatially}
La{\v{c}}evi{\'c},~N.; Starr,~F.~W.; Schr{\o}der,~T.; Glotzer,~S. \emph{The
  Journal of chemical physics} \textbf{2003}, \emph{119}, 7372--7387\relax
\mciteBstWouldAddEndPuncttrue
\mciteSetBstMidEndSepPunct{\mcitedefaultmidpunct}
{\mcitedefaultendpunct}{\mcitedefaultseppunct}\relax
\EndOfBibitem
\bibitem[Karmakar \latin{et~al.}(2009)Karmakar, Dasgupta, and
  Sastry]{karmakar2009growing}
Karmakar,~S.; Dasgupta,~C.; Sastry,~S. \emph{Proceedings of the National
  Academy of Sciences} \textbf{2009}, \emph{106}, 3675--3679\relax
\mciteBstWouldAddEndPuncttrue
\mciteSetBstMidEndSepPunct{\mcitedefaultmidpunct}
{\mcitedefaultendpunct}{\mcitedefaultseppunct}\relax
\EndOfBibitem
\bibitem[Flenner and Szamel(2010)Flenner, and Szamel]{flenner2010dynamic}
Flenner,~E.; Szamel,~G. \emph{Physical review letters} \textbf{2010},
  \emph{105}, 217801\relax
\mciteBstWouldAddEndPuncttrue
\mciteSetBstMidEndSepPunct{\mcitedefaultmidpunct}
{\mcitedefaultendpunct}{\mcitedefaultseppunct}\relax
\EndOfBibitem
\bibitem[Xu \latin{et~al.}(2020)Xu, Douglas, and Xu]{xu2020molecular}
Xu,~W.-S.; Douglas,~J.~F.; Xu,~X. \emph{Macromolecules} \textbf{2020},
  \emph{53}, 4796--4809\relax
\mciteBstWouldAddEndPuncttrue
\mciteSetBstMidEndSepPunct{\mcitedefaultmidpunct}
{\mcitedefaultendpunct}{\mcitedefaultseppunct}\relax
\EndOfBibitem
\bibitem[Nandi \latin{et~al.}(2021)Nandi, Kob, and
  Maitra~Bhattacharyya]{nandi2021connecting}
Nandi,~U.~K.; Kob,~W.; Maitra~Bhattacharyya,~S. \emph{The Journal of Chemical
  Physics} \textbf{2021}, \emph{154}, 094506\relax
\mciteBstWouldAddEndPuncttrue
\mciteSetBstMidEndSepPunct{\mcitedefaultmidpunct}
{\mcitedefaultendpunct}{\mcitedefaultseppunct}\relax
\EndOfBibitem
\bibitem[Xu \latin{et~al.}(2016)Xu, Douglas, and Freed]{xu2016influence}
Xu,~W.-S.; Douglas,~J.~F.; Freed,~K.~F. \emph{Macromolecules} \textbf{2016},
  \emph{49}, 8355--8370\relax
\mciteBstWouldAddEndPuncttrue
\mciteSetBstMidEndSepPunct{\mcitedefaultmidpunct}
{\mcitedefaultendpunct}{\mcitedefaultseppunct}\relax
\EndOfBibitem
\bibitem[Wang \latin{et~al.}(2018)Wang, Xu, Wang, and Guan]{wang2018revealing}
Wang,~L.; Xu,~N.; Wang,~W.; Guan,~P. \emph{Physical review letters}
  \textbf{2018}, \emph{120}, 125502\relax
\mciteBstWouldAddEndPuncttrue
\mciteSetBstMidEndSepPunct{\mcitedefaultmidpunct}
{\mcitedefaultendpunct}{\mcitedefaultseppunct}\relax
\EndOfBibitem
\bibitem[Wang \latin{et~al.}(2019)Wang, Xu, Zhang, and
  Douglas]{wang2019universal}
Wang,~X.; Xu,~W.-S.; Zhang,~H.; Douglas,~J.~F. \emph{The Journal of chemical
  physics} \textbf{2019}, \emph{151}, 184503\relax
\mciteBstWouldAddEndPuncttrue
\mciteSetBstMidEndSepPunct{\mcitedefaultmidpunct}
{\mcitedefaultendpunct}{\mcitedefaultseppunct}\relax
\EndOfBibitem
\bibitem[Zhang \latin{et~al.}(2019)Zhang, Wang, Chremos, and
  Douglas]{zhang2019superionic}
Zhang,~H.; Wang,~X.; Chremos,~A.; Douglas,~J.~F. \emph{The Journal of chemical
  physics} \textbf{2019}, \emph{150}, 174506\relax
\mciteBstWouldAddEndPuncttrue
\mciteSetBstMidEndSepPunct{\mcitedefaultmidpunct}
{\mcitedefaultendpunct}{\mcitedefaultseppunct}\relax
\EndOfBibitem
\bibitem[Zhang \latin{et~al.}(2021)Zhang, Wang, Yu, and
  Douglas]{zhang2021dynamic}
Zhang,~H.; Wang,~X.; Yu,~H.-B.; Douglas,~J.~F. \emph{The European Physical
  Journal E} \textbf{2021}, \emph{44}, 1--30\relax
\mciteBstWouldAddEndPuncttrue
\mciteSetBstMidEndSepPunct{\mcitedefaultmidpunct}
{\mcitedefaultendpunct}{\mcitedefaultseppunct}\relax
\EndOfBibitem
\bibitem[Charbonneau \latin{et~al.}(2012)Charbonneau, Ikeda, Parisi, and
  Zamponi]{charbonneau2012dimensional}
Charbonneau,~P.; Ikeda,~A.; Parisi,~G.; Zamponi,~F. \emph{Proceedings of the
  National Academy of Sciences} \textbf{2012}, \emph{109}, 13939--13943\relax
\mciteBstWouldAddEndPuncttrue
\mciteSetBstMidEndSepPunct{\mcitedefaultmidpunct}
{\mcitedefaultendpunct}{\mcitedefaultseppunct}\relax
\EndOfBibitem
\bibitem[Sastry(2000)]{sastry2000PRL}
Sastry,~S. \emph{Physical Review Letters} \textbf{2000}, \emph{85}, 590\relax
\mciteBstWouldAddEndPuncttrue
\mciteSetBstMidEndSepPunct{\mcitedefaultmidpunct}
{\mcitedefaultendpunct}{\mcitedefaultseppunct}\relax
\EndOfBibitem
\end{mcitethebibliography}



\end{document}